\documentclass[sigconf, screen]{acmart}

\usepackage{mylatexstyle}

\usepackage{amsmath}
\usepackage[ruled,vlined, linesnumbered]{algorithm2e}
\usepackage{graphicx} 
\usepackage{makecell} %
\usepackage{booktabs} %
\usepackage{caption} %
\usepackage{subcaption} %
\usepackage{xcolor}
\usepackage{multirow}

\usepackage{comment} %

\usepackage{tikz} %
\newcommand*\circled[1]{\tikz[baseline=(char.base)]{
            \node[shape=circle,fill=black,text=white,draw,inner sep=0.5pt] (char) {#1};}}

\newcommand{\sysname}[0]{\textit{Heta}} %

\AtBeginDocument{%
  \providecommand\BibTeX{{%
    \normalfont B\kern-0.5em{\scshape i\kern-0.25em b}\kern-0.8em\TeX}}}
    
\renewcommand\footnotetextcopyrightpermission[1]{}
\setcopyright{acmlicensed}
\copyrightyear{2025}
\acmYear{2025}
\acmDOI{XXXXXXX.XXXXXXX}

\acmConference[Conference acronym 'XX]{Make sure to enter the correct
  conference title from your rights confirmation emai}{June 03--05,
  2018}{Woodstock, NY}
\acmISBN{978-1-4503-XXXX-X/18/06}

\begin{document}

\title{\sysname{}: Distributed Training of Heterogeneous Graph Neural Networks}

\author{Yuchen Zhong} 
\affiliation{
    \institution{The University of Hong Kong}
    \city{Hong Kong}
    \country{China}
}
\email{yczhong@cs.hku.hk}

\author{Junwei Su} 
\affiliation{
    \institution{The University of Hong Kong}
    \city{Hong Kong}
    \country{China}
}
\email{junweisu@connect.hku.hk}

\author{Chuan Wu} 
\affiliation{
    \institution{The University of Hong Kong}
    \city{Hong Kong}
    \country{China}
}
\email{cwu@cs.hku.hk}

\author{Minjie Wang} 
\affiliation{
    \institution{AWS Shanghai AI Lab}
    \city{Shanghai}
    \country{China}
}
\email{minjiw@amazon.com}

\begin{abstract}
Heterogeneous Graph Neural Networks (HGNNs) leverage diverse semantic relationships in Heterogeneous Graphs (HetGs) and have demonstrated remarkable learning performance in various applications. However, current distributed GNN training systems often overlook unique characteristics of HetGs, such as varying feature dimensions and the prevalence of missing features among nodes, leading to suboptimal performance or even incompatibility with distributed HGNN training. We introduce \sysname{}, a framework designed to address the communication bottleneck in distributed HGNN training. \sysname{} leverages the inherent structure of HGNNs -- independent relation-specific aggregations for each relation, followed by a cross-relation aggregation -- and advocates for a novel Relation-Aggregation-First computation paradigm. It performs relation-specific aggregations within graph partitions and then exchanges partial aggregations. This design, coupled with a new graph partitioning method that divides a HetG based on its graph schema and HGNN computation dependency, substantially reduces communication overhead. \sysname{} further incorporates an innovative GPU feature caching strategy that accounts for the different cache miss-penalties associated with diverse node types. Comprehensive evaluations of various HGNN models and large heterogeneous graph datasets demonstrate that \sysname{} outperforms state-of-the-art systems like DGL and GraphLearn by up to 5.8$\times$ and 2.3$\times$ in end-to-end epoch time, respectively.
\end{abstract}


\maketitle

\pagestyle{plain}

\setlength{\textfloatsep}{3pt}%
\section{Introduction}

Heterogeneous Graphs (HetGs) are complex network structures that encapsulate diverse semantic relationships between different types of nodes and edges. They are ubiquitous in real-world scenarios: in an academic network~\cite{wang2020microsoft}, nodes represent %
authors, institutions, or papers, and edges denote relationships like authorship, affiliation, or citation; in an e-commerce network~\cite{ni2019justifying}, nodes represent users, products, and categories, with edges indicating interactions such as purchases and views. Heterogeneous Graph Neural Networks (HGNNs) have been designed to learn from HetGs by encoding their rich semantic and structural information~\cite{schlichtkrull2018modeling, busbridge2019relational, hu2020heterogeneous, wang2019heterogeneous}. HGNNs have been shown to outperform homogeneous Graph Neural Networks (GNNs) in graph learning tasks, such as %
recommendation systems~\cite{zheng2021multi}, social network analysis~\cite{gao2022hetinf}, and cybersecurity~\cite{hu2019cash}. %

Training HGNNs on large-scale HetGs with millions of nodes and billions of edges remains challenging~\cite{hu2021ogblsc}. As with distributed GNN training on homogeneous graphs, inter-machine communication for feature fetching is a crucial bottleneck %
in distributed HGNN training due to cross-machine HetG partitioning~\cite{zheng2022distributed}. %
Nonetheless, HetGs exhibit unique characteristics that do not exist in homogeneous graphs, requiring additional system support. %
In a homogeneous graph, nodes typically share a common feature space with uniform feature dimensions~\cite{zheng2022distributed}.
In contrast, HetGs can have significant variations in feature dimensions across different node types; for example, in the Donor dataset~\cite{donor}, the feature dimension varies from 7 to 789 among different node types. %
Some node types in real-world HetGs lack features entirely; for instance, only the paper nodes in the MAG240M dataset~\cite{hu2020open} have features, while others (approximately half of all nodes) do not. %
A prevalent solution in HGNN learning is to assign learnable features to such featureless nodes and update them via backpropagation during HGNN training~\cite{zheng2020distdgl, kalantzi2021position, yin2022nimble, han2022openhgnn}. These learnable features are typically stored in CPU DRAM due to the limited GPU memory, and updating them incurs large overhead due to massive random read/write operations, which deplete DRAM bandwidth~\cite{xie2022fleche} and can account for up to 35\% of the HGNN training time (\S\ref{sec:challenges}).

Most existing distributed GNN systems are designed for homogeneous graphs, %
focusing on communication reduction. Transferring their solutions %
to HGNNs would be either inapplicable or suboptimal. %
First, techniques like $P^3$\cite{gandhi2021p3}, which partition node features along the feature dimension, assume uniform feature dimension sizes across all nodes -- a method inapplicable to HetGs with varying feature dimensions. Next, prior works propose caching features of frequently accessed nodes on GPUs to reduce feature fetching%
~\cite{jia2020roc, lin2020pagraph, yang2022gnnlab, kaler2023communication, sun2023legion}. 
Their caching methods only consider the ``hotness'' of nodes and overlook variations in the cache ``miss penalty'' of each node type, such as varying feature dimensions and whether features are learnable. Feature dimensions affect the retrieval cost of data while missing learnable features increase the penalty due to additional updates (\S\ref{sec:gpu_cache}). Consequently, these caching methods are suboptimal when dealing with HGNNs and HetGs. 

Distributed GNN training frameworks such as DGL~\cite{zheng2022distributed} and GraphLearn~\cite{graphlearn} technically support distributed training of HGNNs, but they achieve suboptimal performance by overlooking the unique characteristics of HetGs and HGNNs. These systems train HGNNs in a manner similar to homogeneous GNNs, adhering to the \textit{vanilla execution model} %
using data parallelism and edge-cut partitioning. %
DGL converts a HetG topology (which includes multiple adjacency matrices for different relations) into a single adjacency matrix and then applies minimizing-edge-cut algorithms like METIS~\cite{metis1998} for graph partitioning. %
GraphLearn uses random edge-cut partitioning for nodes of each node type. The rest of the HGNN training process remains the same as homogeneous GNNs. %
While these partitioning methods aim to minimize cross-partition edges, 
they are suboptimal for HGNNs as they do not fully leverage the inherent computational structure of HGNNs (%
\S\ref{sec:computation_model}). As a result, these systems still suffer from significant communication overhead for exchanging features of boundary nodes (i.e., nodes with edges crossing different partitions)~\cite{gandhi2021p3, liu2021bgl} and can be further optimized for HetGs.

\begin{figure}[t]
    \centering
    \includegraphics[width=0.85\columnwidth]{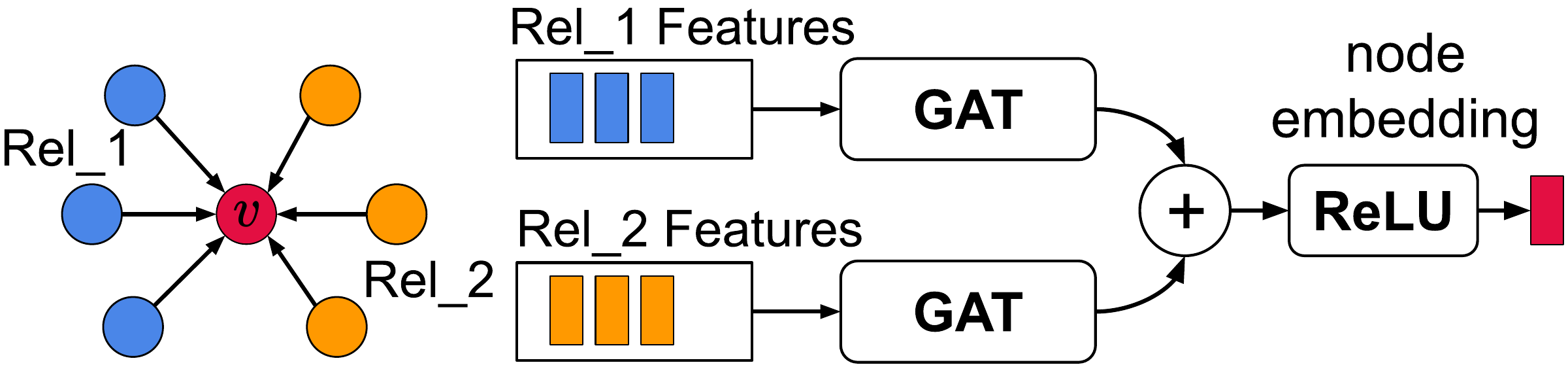}
    \vspace{-3mm}
    \caption{Illustration of R-GAT model architecture.} 
    \label{fig:embedding_model}
\end{figure}

To efficiently address the communication bottleneck in distributed HGNN training, we propose \sysname{}, a novel distributed framework with a co-designed communication reduction paradigm (\S\ref{sec:computation_model}) and graph partitioning (\S\ref{sec:meta-partitioning}). The key insight behind \sysname{} is that the aggregation of a HGNN can be decomposed into multiple relation-specific aggregations on mono-relation subgraphs, each corresponding to a single relation %
in the HetG. Figure~\ref{fig:embedding_model} depicts a representative HGNN model, R-GAT~\cite{busbridge2019relational}, where two GAT modules~\cite{busbridge2019relational} aggregate features from different mono-relation subgraphs, and their outputs are combined and activated by ReLU to generate node embeddings.
This decomposition is applicable to most HGNNs (\S\ref{sec:hgnn}) %
and has not been fully leveraged by current distributed GNN systems.

Based on this insight, we propose a novel HGNN computation method called the \textit{Relation-Aggregation-First (RAF) paradigm} to reduce inter-machine communication overhead (\S\ref{sec:computation_model}).
The input HetG is partitioned into multiple partitions based on relations, each containing one or more complete mono-relation subgraphs decomposed from the HetG.
RAF conducts relation-specific aggregations within each partition %
and subsequently exchanges the partial aggregations among partitions to compute the final node embeddings using model parallelism. During backpropagation, the gradients of the partial aggregations are exchanged across partitions. RAF substantially reduces communication overhead by transmitting only intermediate aggregations and gradients, thereby entirely avoiding the need to move features.  %
The smaller hidden dimension of GNN models further decreases the size of data transmission~\cite{gandhi2021p3}.  %
We %
prove that the communication complexity under RAF depends on the maximum number of boundary nodes among the partitions, which is lower than the number of cross-partition edges in the vanilla execution model %
used by existing systems. %

According to our communication complexity analysis, our HetG partitioning aims to minimize the maximal number of boundary nodes among partitions. %
However, solving the optimal graph partition problem is computationally infeasible due to its NP-hard nature~\cite{metis1998}.
Drawing on the concept of a metagraph 
(a high-level abstraction of a HetG), we introduce an efficient new graph partitioning algorithm called \textit{meta-partitioning} (\S\ref{sec:meta-partitioning}). 
Meta-partitioning assigns strongly connected relations in the metagraph to the same partition by analyzing the HGNN computation dependency graph. It balances the number of nodes/edges and target nodes among partitions to ensure balanced workloads 
and restricts boundary nodes to %
target nodes to keep constant communication complexity. %

We further devise a novel GPU caching strategy to reduce the overhead of updating learnable features (\S\ref{sec:gpu_cache}). We find that \textit{miss-penalty ratios} (i.e., the time penalty per unit cache size incurred upon a cache miss) vary significantly across node types, influenced by feature dimensions and whether features are learnable (Figure~\ref{fig:miss_penalty_ratio}). Therefore, we propose a cache size allocation strategy based on node hotness and miss-penalty ratios. Additionally, we extend caching to include mutable learnable features and optimizer states, promoting a non-redundant design for consistency.

We implement \sysname{} on DGL~\cite{zheng2022distributed} and PyTorch~\cite{paszke2019pytorch}. Comprehensive evaluations (\S\ref{sec:evaluation}) demonstrate the effectiveness and efficiency of our approaches, surpassing state-of-the-art systems DGL and GraphLearn by up to 5.8$\times$ and 2.3$\times$ in end-to-end epoch time, respectively, without any loss in accuracy when training representative HGNN models on large heterogeneous graphs. Moreover, \sysname{}'s meta-partitioning demonstrates superior efficiency in terms of both time and memory footprint compared to state-of-the-art graph partitioning methods such as METIS. %

\section{Background and Motivation}
\label{sec:background_and_motivation}

\subsection{Heterogeneous Graph Neural Networks}
\label{sec:hgnn}

\begin{figure}[t]
    \centering
    \includegraphics[width=0.9\columnwidth]{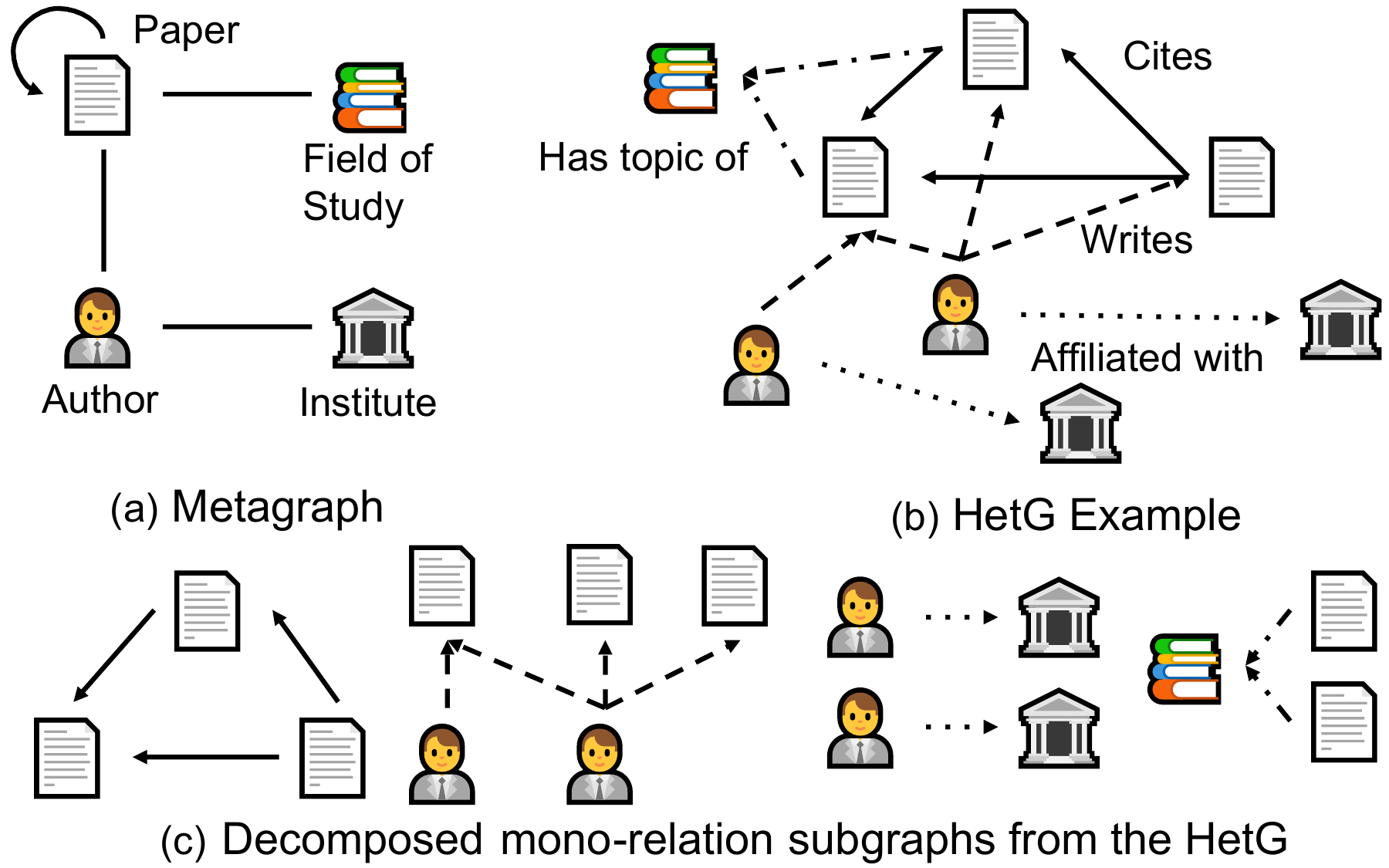}
    \vspace{-3mm}
    \caption{Metagraph and decomposed mono-relation subgraphs of the ogbn-mag dataset~\cite{hu2020open}.}
    \label{fig:metagraph}
\end{figure}

\noindent\textbf{Heterogeneous Graphs (HetGs).}
A %
HetG is denoted by $G = (V, E, A, R)$, where each node $v\in V$ and each edge $e \in E$ are associated with type mapping functions $\tau(v): V\rightarrow A$ and $\phi(e): E\rightarrow R$, respectively. The set $A$ represents the node types, while $R$ represents the edge types that connect the nodes. 
A relation $r$ on edge $e=(u,v)$ %
is defined as a triple $(\tau(u), \phi(e), \tau(v))$, and $\mathcal{R}$ denotes the set of all such relations. The reverse relation $r^{-1}$ for $r$ is $(\tau(v), \bar{\phi}(e), \tau(u))$, where $\bar{\phi}(e)$ represents the inverse connection type on edge $e$. %
 A mono-relation subgraph can be defined for each relation, containing nodes of types $\tau(u)$ and $\tau(v)$ and edges of type $\phi(e)$.  
 The HetG can be decomposed into a collection of mono-relation subgraphs %
 based on the relations. %
The structure of a HetG can be further described by a metagraph $M = (A, R)$ and the relationships between nodes captured by metapaths, which are sequences characterized by the types of nodes and edges on the paths connecting the nodes, representing various semantic relations between nodes. %
Some nodes are associated with dense feature vectors, and each node type may correspond to a feature vector of a different shape; some node types may lack features entirely. %
Typically, one node type is of particular interest and is associated with labels, referred to as the target node type in the HetG. They are also referred to as training nodes.  

Figure~\ref{fig:metagraph} illustrates a concrete HetG and the metagraph %
from the ogbn-mag dataset~\cite{hu2020open}, 
consisting of four node types and four relations (plus three reverse relations). Figure~\ref{fig:metagraph} (c) shows the mono-relation subgraphs decomposed from the HetG. %
Only the node type ``paper'' is associated with node features, while other node types do not. The target node type %
is ``paper'' associated with venue labels, and the task of this dataset is to predict the venue of a given paper.

\vspace{1mm}
\noindent\textbf{Heterogeneous Graph Neural Networks (HGNNs).}
Given a HetG and node $v$, the HGNN computation to 
generate node embedding $\mathbf{h}^{(l)}_v$ at its layer $l$ is: %
{\small
\begin{equation} \label{eq:hgnn}
\begin{gathered}
    \mathbf{h}^{(l)}_{v, r} = \color{blue}\text{AGG}^{(l)}_{r}\color{black}\left(\left\{\mathbf{h}^{(l-1)}_{u}, u\in N_{\mathcolor{blue}r}(v)\right\} \right) \\
    \mathbf{h}^{(l)}_{v} = \color{blue}\text{AGG}^{(l)}_{\text{all}}\color{black}\left(\left\{\mathbf{h}^{(l)}_{v, r}, \color{blue}r \in \mathcal{R} \color{black}\right\} \right)
\end{gathered}
\end{equation}
}

\noindent where $\text{AGG}_{r}^{(l)}$ is a relation-specific aggregation function for relation $r$ (typically a GNN like GCN~\cite{kipf2017semisupervised}), $\mathbf{h}^{(l)}_{v, r}$ is the partial embedding of relation $r$ using neighbors $N_r(v)$ (i.e., the neighbors of node $v$ under relation $r$), and $\text{AGG}^{(l)}_{\text{all}}$ is a cross-relation aggregation function (typically a reduce function like summation). $\mathbf{h}^{0}_v$ is initialized using the feature vector of node $v$ or its learnable feature vector $\mathbf{w}_v$ 
(if node features are not available)~\cite{zheng2022distributed}.

Eq.~(\ref{eq:hgnn}) suggests that the aggregation of a HGNN can be decomposed into individual relation-specific aggregations for each relation $r\in \mathcal{R}$, followed by a %
cross-relation aggregation. Two representative HGNNs, R-GCN~\cite{schlichtkrull2018modeling} and R-GAT~\cite{busbridge2019relational}, %
apply GCN~\cite{kipf2017semisupervised} and GAT~\cite{velickovic2018graph}, respectively, for relation-specific aggregation, and then sum the embeddings.
These models have a set of weight matrices for each relation. %
HGT uses weight matrices for each node and edge type instead of each relation~\cite{hu2020heterogeneous}, allowing it to perform relation-specific aggregation using the parameters associated with corresponding node and edge types.

\subsection{Distributed Training of HGNNs}
\label{sec:existing_systems}

\begin{figure}[t]
    \centering
    \includegraphics[width=0.95\columnwidth]{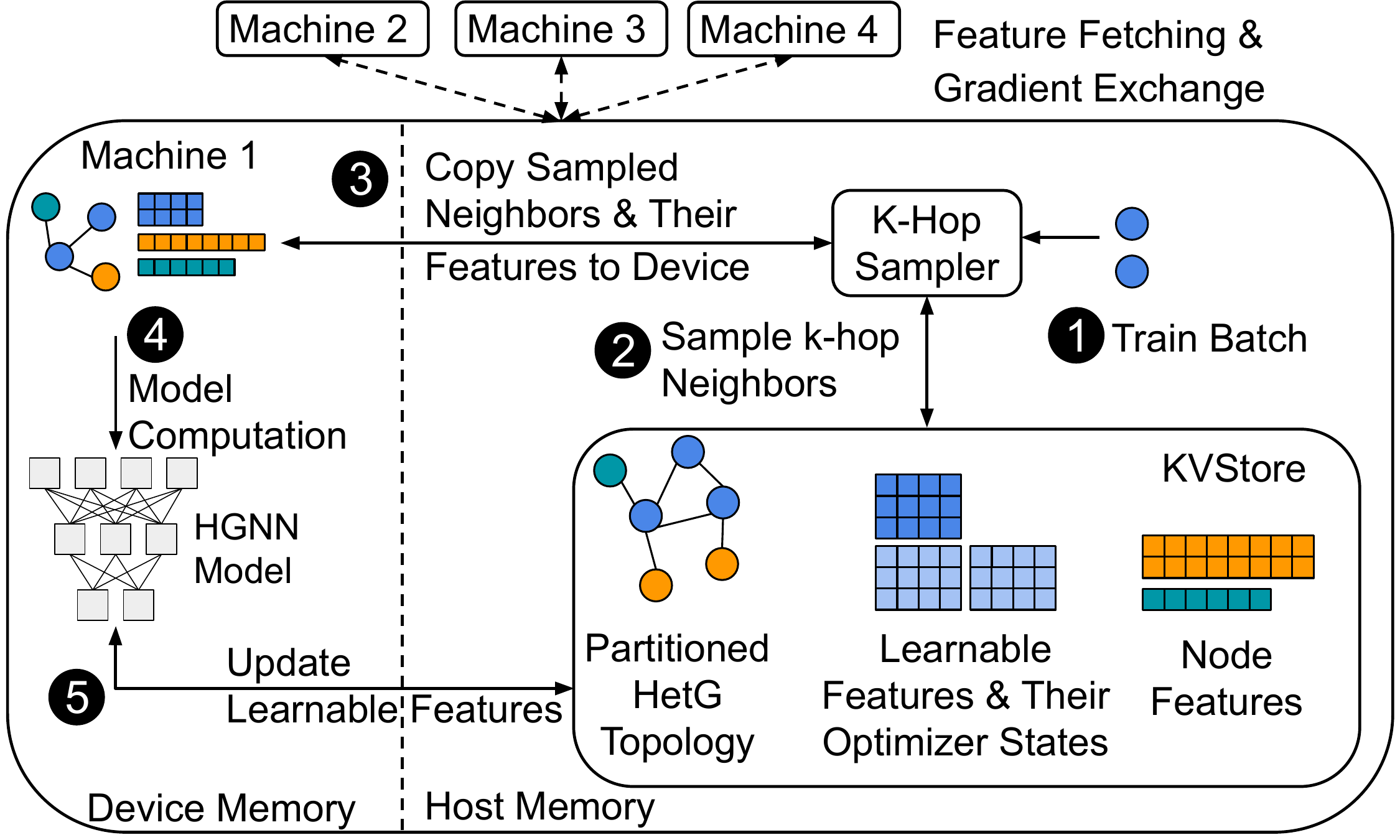}
    \vspace{-3mm}
    \caption{Vanilla execution model of existing distributed GNN training systems that support HGNN training.} 
    \label{fig:existing_system}
\end{figure}

Existing distributed GNN training systems such as DGL~\cite{zheng2022distributed} and GraphLearn~\cite{graphlearn} %
use edge-cut graph partitioning and data-parallel training, which we refer to as the \textit{vanilla execution model}. %
DGL supports assigning learnable features for featureless nodes, while GraphLearn does not. Some optimizers in GNN training, such as the Adam optimizer~\cite{kingma2014adam}, store optimizer states (such as moments and variances) that are the same size as the learnable features. Node features, learnable features, and their optimizer states are partitioned according to the associated nodes.

The distributed data-parallel training process of existing GNN systems involves several steps (Figure~\ref{fig:existing_system}). Each worker receives a mini-batch of training nodes (\circled{1}) and passes them to the %
sampler, which samples a fixed number of $k$-hop neighbors (\circled{2})%
~\cite{hamilton2017inductive}. %
The node features or learnable features of the sampled neighbors are fetched from the key-value store (KVStore) in host memory. If the required data are unavailable locally, network communication is needed to retrieve them from other machines. The sampled neighbors and associated features are then copied to GPU memory (\circled{3}) for HGNN model computation. Gradients are synchronized among the GPUs for model parameter updates. Finally, %
optimizer states are fetched from the local KVStore to GPU memory; both the optimizer states and learnable features are updated according to the optimizer and then written back to the local KVStore (\circled{5}).

\begin{figure}[t]
    \centering
    \includegraphics[width=\columnwidth]{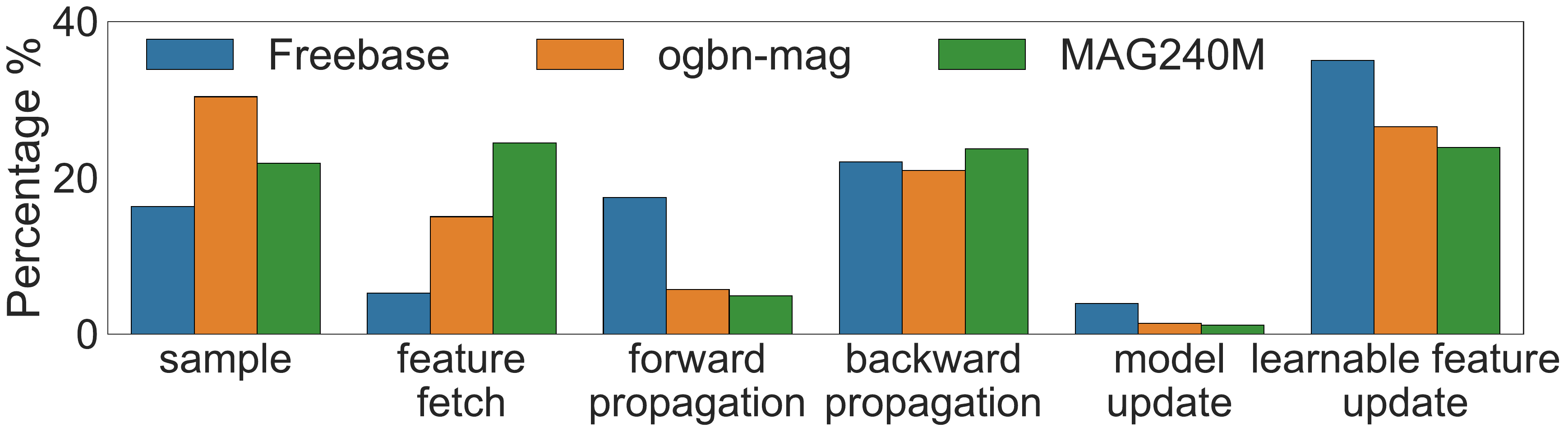}
    \vspace{-6mm}
    \caption{Percentage of the epoch time spent on each stage: training R-GCN on three datasets with DGL.}
    \label{fig:distdgl_profile}
\end{figure}

\subsection{Opportunities \& Challenges}
\label{sec:challenges}

\begin{figure*}[t]
    \centering
    \includegraphics[width=0.85\textwidth]{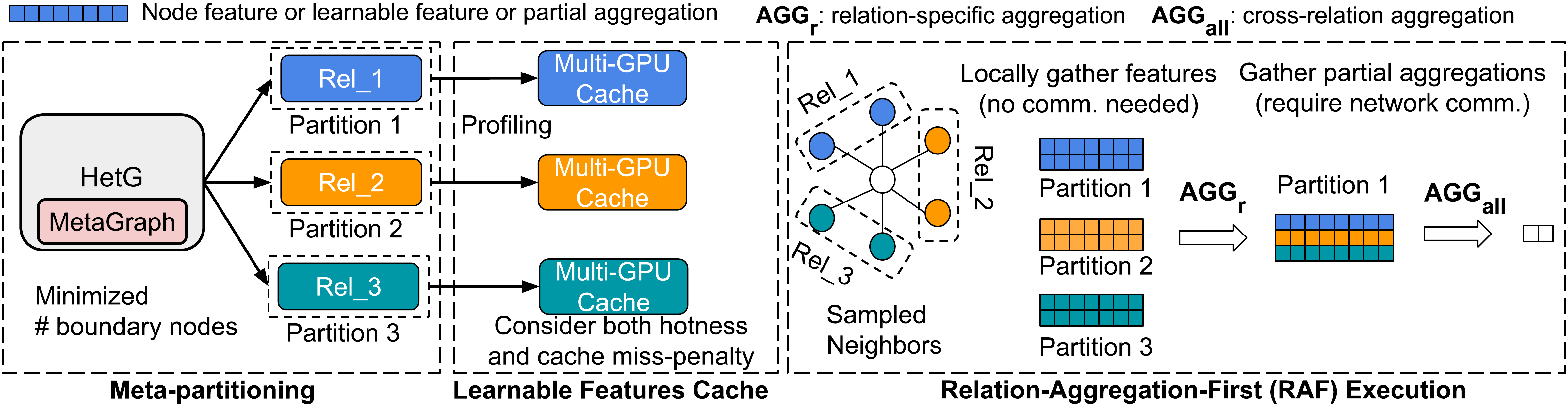}
    \vspace{-3mm}
    \caption{\sysname{}'s design overview.}
    \vspace{-3mm}
    \label{fig:overview}
\end{figure*}

\noindent\textbf{Challenge 1: Communication Bottleneck.}
As noted in prior works~\cite{gandhi2021p3, liu2021bgl}, distributed GNN training suffers from inter-machine communication bottlenecks. 
Existing communication reduction methods designed for homogeneous GNNs are insufficient for HGNNs. %
$P^3$~\cite{gandhi2021p3} %
partitions node features by feature dimension, assuming uniform feature size across nodes. %
Other methods, such as dropping boundary nodes~\cite{wan2022bns}, allowing node embedding 
staleness~\cite{wan2021pipegcn, fey2021gnnautoscale}, and feature quantization~\cite{wan2023adaptive}, %
reduce communication volumes but alter model computational equivalence. %

\vspace{1mm}
\noindent\textbf{Opportunity 1: Relation-aware Model Parallelism.}
Eq. (\ref{eq:hgnn}) indicates that HGNN computation can be split among workers according to the relation type $r$, allowing each worker to aggregate embeddings within a relation before communicating with other workers, thus reducing communication overhead. This amounts to model parallelism for HGNN training. %
We explore this insight and propose a Relation-Aggregation-First (RAF) computation paradigm that preserves model mathematical equivalence (\S\ref{sec:computation_model}).

\vspace{1mm}
\noindent\textbf{Challenge 2: Lack of Effective Partitioning Strategy for HetGs.}
The most widely adopted graph partitioning scheme in distributed GNN systems is edge-cut partitioning, particularly METIS~\cite{metis1998, zheng2020distdgl, wan2022bns, sun2023legion}. %
These %
methods mainly focus on homogeneous graphs %
and overlook %
the computational dependency of HGNNs (i.e., the aggregation path defined by HetG multi-hop sampling or metapaths). %
For example, the ``Paper-Author-Paper'' metapath aggregates data through shared authors. However, METIS-type partitioning treats all edges as equivalent, ignoring node types and specific paths required for HGNN computation.
                                            
\vspace{1mm}            
\noindent\textbf{Opportunity 2: Relation-aware Partitioning.}
By considering HGNN computation dependency, relation-aware partitioning can ensure that connected 
node types  %
and their connections are more likely to reside within the same partition. 
This requires designing new partitioning objectives and methods. We propose meta-partitioning, which partitions a HetG based on its metagraph and leverages HGNN computation dependency (\S\ref{sec:meta-partitioning}).

\vspace{1mm}
\noindent\textbf{Challenge 3: Large Overhead of Learnable Features Update.} We further identify
a unique problem of HGNN training -- the large overhead of updating learnable features. %
Figure \ref{fig:distdgl_profile} shows a training time breakdown when training R-GCN on the Freebase~\cite{freebase:datadumps}, ogbn-mag~\cite{hu2020open}, and MAG240M~\cite{hu2021ogblsc} datasets with DGL and METIS partitioning on two Amazon EC2 g4dn.metal instances. The Freebase dataset includes no node features, and learnable features are used. Only paper nodes have features in the other two datasets, and other node types do not. Updating learnable features accounts for 24\% to 35\% of the epoch time, primarily due to the substantial random DRAM read/write operations for learnable features and optimizer states from/to the host memory~\cite{xie2022fleche}.

\vspace{1mm}
\noindent\textbf{Opportunity 3: GPU Cache for Learnable Features.}
Previous works have proposed caching frequently accessed read-only node features %
on GPUs to alleviate %
DRAM I/O operations~\cite{lin2020pagraph, yang2022gnnlab, sun2023legion}. They %
do not consider %
the heterogeneity of feature dimensions nor learnable features in HetGs. %
We propose a novel GPU cache that incorporates miss-penalty-aware cache size allocation to address the heterogeneity of feature dimensions and learnable features. Also, we propose a non-replicative cache design that caches learnable features and their optimizer states (\S\ref{sec:gpu_cache}).

\section{System Overview}
\label{sec:overview}

\sysname{} is a distributed framework %
for efficient distributed HGNN training on HetGs. An overview of the design %
is presented in Figure~\ref{fig:overview}. %
\sysname{} employs meta-partitioning (\S \ref{sec:meta-partitioning}), a relation-aware partitioning scheme that decomposes the heterogeneous graph into balanced partitions based on its metagraph and HGNN computation dependency. Each partition comprises one or more complete mono-relation subgraphs (\S\ref{sec:hgnn}). Before training, %
\sysname{} profiles the access frequency of nodes and miss-penalty ratios for different node types, calculates the cache size for each node type accordingly, and initializes the cache based on these access frequencies. 
\sysname{} utilizes mini-batch sampling-based, iterative GNN training and introduces a novel Relation-Aggregation-First (RAF) paradigm. This approach aggregates the partial results to generate node embeddings over the graph partitions without moving features (\S\ref{sec:computation_model}). After backpropagation in each training iteration, the model parameters and the learnable features are updated.

\section{RAF HGNN Computation%
}\label{sec:computation_model}

\begin{algorithm}[!th]
\DontPrintSemicolon
\SetAlgoLined
\SetKwBlock{DoParallel}{for each worker $i$ do in parallel}{end}
\SetKwBlock{DoWorker}{on worker $w_d$ do}{end}

\KwIn{Partitioned heterogeneous graph $G$ based on relations $\{G_1, G_2, \ldots, G_{p}\}$; set of target nodes $V_{\text{target}}$; the $k$-layer HGNN; designated worker $w_d$}
\KwOut{Updated model parameters $\theta$ of the HGNN}
\BlankLine

\ForEach{batch $B \subseteq V_{\text{target}}$}{
  $S_B^L \gets$ Sample $k$-hop neighbors for nodes in $B$\;
  \For{$l=1$ \KwTo $k$}{
    \tcc{Relation-specific aggregation}
    \DoParallel{
      $\mathbf{h}^{(l)}_i \gets$ Aggregate neighbor features/embeddings of $S_B^l$ for the $i$-th partition's relations\;
      Send partial aggregations $\mathbf{h}^{(l)}_i$ to 
      worker $w_d$\;
    }
    \DoWorker{
      \tcc{Cross-relation aggregation}
      $\mathbf{h}^{(l)} \gets \text{AGG}^{(l)}_{\text{all}}(\{\mathbf{h}^{(l)}_i \})$\;
      \If{$l = k$}{
        Compute loss and backpropagate\;
        Send gradients $\nabla \mathbf{h}^{(l)}_i$ to workers\;
      }
    }
    \DoParallel{
 
      \If{$l = k$}{
        Backpropagate with 
        $\nabla \mathbf{h}^{(l)}_i$ and
        update local model parameters $\theta_i$\;
      }
    }
  }
}
\caption{RAF Execution Paradigm}
\label{alg:raf_paradigm}
\end{algorithm}

We present the \textit{Relation-Aggregation-First (RAF)} paradigm in Algorithm~\ref{alg:raf_paradigm}, which exploits the computational structure of HGNNs and provably reduces communication complexity compared to the vanilla execution model.

\vspace{1mm}
\noindent\textbf{RAF Paradigm.}
Each worker maintains a partition of the input HetG, %
holds only the model parameters for the relations %
in its partition, and performs relation-specific aggregations (referred to as partial aggregation) and model updates locally. One %
randomly assigned worker, i.e., the designated worker, handles cross-relation aggregation.
Specifically, each worker aggregates features or the %
previous layer's embeddings of the sampled neighborhood of the target nodes %
for its local relations %
(line 5, Algorithm~\ref{alg:raf_paradigm}) 
without incurring any network communication. %
The designated worker then collects all partial aggregations from other workers (line 6, involving network transmissions), %
combines embeddings from different relations (line 9), and performs loss computation and backpropagation (line 11). Gradients of partial aggregations are sent back to the corresponding workers (line 12), which then compute gradients and update their model parameters %
(lines 15-19). The learnable features of nodes in a worker's partition are part of the model parameters on that worker and are updated together. %

\vspace{1mm}
\noindent\textbf{Communication Reduction.}
Our RAF paradigm achieves significant communication reduction in two ways. %
{\em First}, it dramatically reduces the number of messages exchanged. In the vanilla execution model of existing GNN systems, features of sampled $k$-hop neighbors must be fetched.
RAF eliminates inter-machine transfers for feature fetching and only requires network communication of partial aggregations and gradients from the inner hops of a target node's neighborhood, as outer hop information (i.e., features) is already locally stored on each partition. Outer-hop features are local because each partition contains complete mono-relation subgraphs, allowing local aggregation, while inner-hop partial aggregations may need communication if they span different partitions.
{\em Second}, the communication messages themselves are smaller. %
Features are usually of high dimensions, %
while partial aggregations and their gradients %
correspond to the dimensions of the model's hidden layers, %
which are generally smaller. 
Consider training a 2-layer R-GCN model~\cite{schlichtkrull2018modeling} with a hidden dimension 64 and fanouts \{25, 20\} on the MAG240M dataset~\cite{hu2021ogblsc}. 
Paper nodes have a feature dimension of 768, and other node types have a learnable feature dimension of 64, all in float16 format. Suppose a batch of 1024 training nodes is sampled and the graph partitioned onto two machines using METIS~\cite{metis1998}. In total 289,986 neighbor nodes are sampled in layer 0 (2-hop), with 66,772 stored on remote machines. Fetching features of these remote nodes and their topology with the vanilla execution model requires transmitting 92.3 MB of data. RAF, however, only requires exchanging partial aggregations and their gradients of 14,568 nodes in layer 1 (1-hop) and 1024 nodes in layer 2 (input), totaling 8.0 MB.

With our meta-partitioning (\S\ref{sec:meta-partitioning}), we can further reduce the communication overhead by combining aggregation messages of multiple relations within the same partition before sending them to the designated worker. %
This approach is highly effective in multi-hop or metapath-based sampling with hierarchical aggregations. %
In the above example, %
only partial aggregations and their gradients of 1024 nodes in layer 2 need to be exchanged, not those in layer 1, as meta-partitioning limits cross-partition dependencies (i.e., boundary nodes) to the target nodes (i.e., ``paper'' nodes),  %
reducing the communication volume to just 0.5 MB.

We demonstrate the mathematical equivalence of RAF with the vanilla execution model in achieving the same HGNN model, analyze its communication complexity, and prove its communication reduction. Detailed proof of the propositions can be found in the supplementary material. 

\begin{proposition}[Mathematical Equivalence]\label{prop:equivalence} Let $\mathbf{h}_v^{(\mathrm{vanilla})}$ and $\mathbf{h}_v^{(\mathrm{RAF})}$ be the embedding of a target node $v$ obtained with the vanilla execution model and the RAF paradigm, respectively. It holds that
$\mathbf{h}_v^{(\mathrm{vanilla})} = \mathbf{h}_v^{(\mathrm{RAF})}$.
\end{proposition}

\noindent This asserts that RAF yields the same representation for any node as the vanilla execution model, %
validating RAF's computation paradigm. Next,  we analyze communication complexity for a HetG divided into two partitions, $G_1$ and $G_2$. The results can be readily extended to multiple partitions. %

\begin{proposition}[Communication Complexity]\label{prop:communication_complexity}
With RAF, the number of communication messages from the worker holding partition $G_1$ to the worker holding partition $G_2$ is proportional to the number of boundary nodes in $G_1$ (nodes with neighbors in the other partition), i.e., $\Theta(|\mathrm{B}(G_1)|)$. The communication complexity in the reverse direction is $\Theta(|\mathrm{B}(G_2)|)$. $\mathrm{B}(G_i)$ denotes the boundary nodes of partition $G_i$. 
\end{proposition}

\noindent The communication complexity under the vanilla execution model scales with the number of cross-partition edges, %
$\mathrm{E}(G_1,G_2)$. The following proposition gives that the number of boundary nodes is always no larger than the count of cross-partition edges.

\begin{proposition}[Communication Reduction]\label{prop:communication_complexity2}
$$\max \{|\mathrm{B}(G_1)|,|\mathrm{B}(G_2)|\} \leq \mathrm{E}(G_1,G_2).$$
\end{proposition}

\noindent Therefore, RAF generally incurs lower communication overhead than the vanilla execution model. %

\section{Meta-Partitioning of HetG}
\label{sec:meta-partitioning}

\begin{figure*}[!th]
    \centering
    \includegraphics[width=0.9\textwidth]{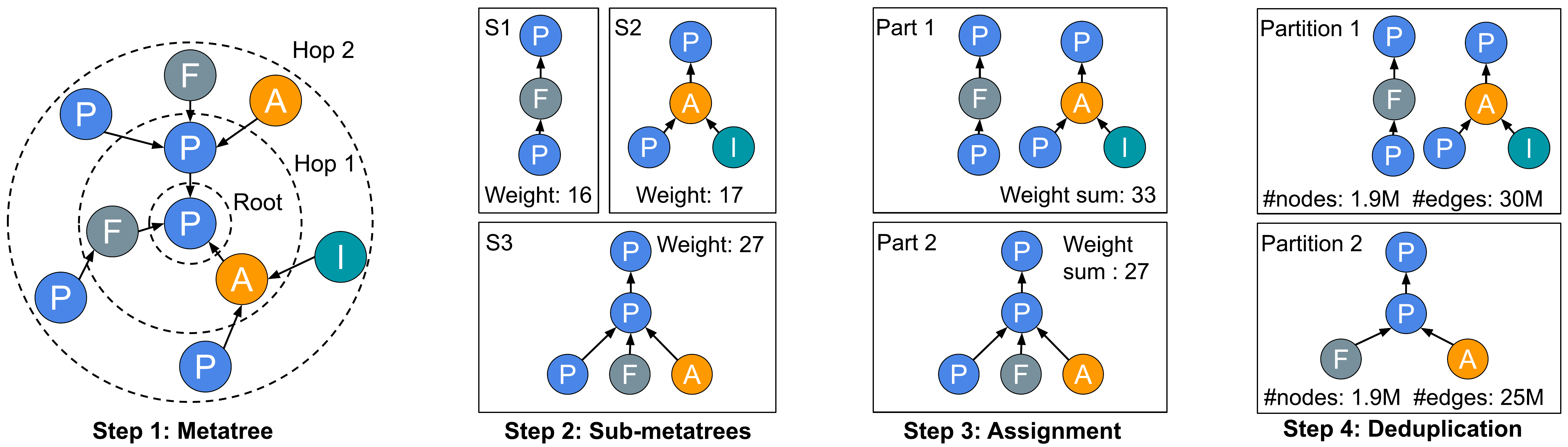}
    \vspace{-2mm}
    \caption{Workflow of meta-partitioning on the ogbn-mag dataset. %
    `P' - paper, `A' - author, `F' - field of study, `I' - institute.} %
    \vspace{-3mm}
    \label{fig:meta_partitioning_workflow}
\end{figure*}

Our analysis in \S\ref{sec:computation_model} shows that communication complexity under RAF depends on the maximal number of boundary nodes among the partitions. We design a HetG partitioning scheme that minimizes this maximal number of boundary nodes. %
Suppose there are $p$ machines used for HGNN training. The $p$-way graph partitioning problem is:
{\small
\begin{equation}\label{eq:graph_partition}
\begin{split}
\text{Minimize} \quad & \max \left\{|\mathrm{B}(G_1)|, |\mathrm{B}(G_2)|, \ldots, |\mathrm{B}(G_p)|\right\} \\
\text{subject to} \quad & G \subseteq G_1 \bigcup G_2 \bigcup \ldots \bigcup G_p, \\
& \text{with balanced partitions.}
\end{split}
\end{equation}
}

\noindent The optimal graph partition problem in %
(\ref{eq:graph_partition}) %
is an NP-hard problem~\cite{metis1998}. We propose an efficient %
graph partitioning heuristic, \textit{meta-partitioning}, %
to partition a HetG according to its metagraph (\S\ref{sec:hgnn}). %
This approach aims to minimize the maximal number of boundary nodes among the partitions and ensures a balanced distribution of nodes/edges and target nodes across partitions to ensure workload balance~\cite{liu2021bgl}. The input to %
meta-partitioning is a weighted metagraph of the heterogeneous graph, where the weights of vertices (links) are the number of nodes for the node type (the number of edges for the edge type) in the HetG.\footnote{To avoid confusion, we use the terms ``vertex'' and ``link'' when referring to the metagraph, and ``node'' and ``edge'' when referring to the HetG.}

\vspace{1mm}
\noindent\textbf{Rationale.}
A naive approach is to randomly assign relations to partitions without considering the unique relation access pattern of HGNN computation or the locality of relations. %
We leverage the computation dependency %
of HGNNs.
In a HetG dataset, training nodes are generally of a specific node type, i.e., the target node type (\S\ref{sec:hgnn}). For a node $u$ of type $\tau(u)$, the set of valid node types that can be sampled in the next hop is given by $\{ \tau(v) | (\tau(u), \phi(e), \tau(v)) \in \mathcal{R} \text{ for some } e = (u, v) \in E \}$. Considering the metagraph, where vertices and links %
represent node and edge types in the HetG, %
for neighborhood sampling,  
we can only start from a specific vertex %
(i.e., target node type) and traverse the metagraph by following valid edge types (links in the metagraph). %
Based on this observation, we derive a \textit{metatree} on the metagraph, rooted at the target node type, representing the computation dependency of HGNNs. An example %
metatree of the ogbn-mag dataset~\cite{hu2020open} with two-hop neighborhood sampling is shown in Figure~\ref{fig:meta_partitioning_workflow} (Step 1).  
Given the metatree derived from the metagraph, we can assign strongly connected relations (i.e., those frequently accessed together during HGNN computation) in the metatree to the same partition and minimize cross-partition dependencies (i.e., boundary nodes).

\begin{algorithm}[t]
\DontPrintSemicolon
\SetAlgoLined
\KwIn{Heterogeneous graph $G$, weighted metagraph $M$, target node type $root$, number of partitions $p$, number of HGNN layers $k$, (optional) user-defined metapaths}
\KwOut{Balanced partitions with minimized boundary nodes}
\eIf{metapaths provided}{
    Construct a metatree $T$ with metapaths\;
}{
    Construct a metatree $T$ with $k$-depth BFS from $root$ in $M$\;
}

\ForEach{child $c$ of $root$ in $T$}{
    Construct a sub-metatree $S_c$ starting from $root$ and containing $c$ and its descendants\;
    Compute weight $w_c$ of $S_c$ as the sum of the weights of vertices and links in $S_c$\;
}
Sort sub-metatrees in descending order of weights\;
\textbf{Initialize} a list of $p$ empty partitions $P = \{\emptyset, \ldots,\emptyset\}$\;
\textbf{Initialize} an array of $p$ sums $sums = [0, 0, \ldots, 0]$%
\;
\ForEach{sub-metatree $S$ in the sorted list of sub-metatrees}{
    Find index $i$ of the partition with the smallest sum in $sums$\;
    Add $S$ to the $i$-th partition in $P$\;
    Add the weight of $S$ to $sums[i]$\;
}
\ForEach{partition $P_i$ in $P$}{
    Deduplicate relations and node features in $P_i$\;%
    Construct HetG partition based on $G$ and $P_i$\;
}
\Return{$P$}
\caption{Meta-Partitioning of HetG}
\label{alg:meta-partitioning}
\end{algorithm}

\vspace{1mm}
\noindent\textbf{Algorithm.} %
The meta-partitioning procedure (Algorithm~\ref{alg:meta-partitioning}) consists of four %
steps. %

\vspace{1mm}
\noindent\textit{Step 1: Build metatree.} 
We use Breadth-First Search (BFS) to construct the metatree, starting from the target node type in the metagraph (line 4%
). BFS explores the graph by expanding the nearest neighbors before going to farther nodes, aligning with how multi-hop sampling works. %
For $k$-hop sampling, the metatree is built with $k$-depth BFS. 
Alternatively, the user can provide a list of metapaths, %
and the algorithm constructs a metatree $T$ using these metapaths starting from the $root$ node %
(line 2).
The metatree of ogbn-mag in Figure \ref{fig:meta_partitioning_workflow} is built by performing a 2-depth BFS starting from the target node type ``paper'' (P) in the metagraph depicted in Figure \ref{fig:metagraph}. %

\vspace{1mm}
\noindent\textit{Step 2: Split metatree.} %
We split the metatree into sub-metatrees to create HetG partitions. %
Each sub-metatree, denoted as $S_c$, corresponds to a child $c$ of the root vertex, %
including the root vertex, the child vertex $c$, and all its descendant vertices (lines 6-7). Each sub-metatree is assigned to a HetG partition, as well as relations in the sub-metatree. %
Since each sub-metatree contains the root vertex of the metatree, each HetG partition contains all nodes of the target node type. %
This setup restricts the boundary nodes of partitions to the target nodes (as no other nodes in each partition are related to nodes in other partitions) %
and ensures HGNN aggregations defined by the aggregation path (i.e., sub-metatrees in a partition) can be performed within each partition.   %
It also helps balance the load since all partitions contain all target nodes. Thus, the maximal number of boundary nodes is upper bound by the number of target nodes.

The weight $w_c$ of each sub-metatree $S_c$ is computed by summing the weights of its vertices and links (line 8) and %
used for balancing HetG partitioning. Balancing the number of nodes among partitions is crucial, as 
retrieving node features from the host memory to the GPU for processing and learnable feature updates %
incurs substantial DRAM accesses and host-device data copies. %
Balancing the number of edges %
is equally important, impacting graph sampling operations.  

In Figure~\ref{fig:meta_partitioning_workflow} (Step 2), the metatree is partitioned into three sub-metatrees, $S_1$, $S_2$, and $S_3$, as the root vertex has three children. %
The weights of $S_1$, $S_2$, and $S_3$ are 16, 17, and 27 million, respectively. %

\vspace{1mm}
\noindent\textit{Step 3: Assignment.}  Assigning the sub-metatrees to $p$ partitions %
is a $p$-way number partitioning problem~\cite{graham1969bounds} that divides the weights of sub-metatrees into $p$ sets to minimize the largest sum of weights among the sets, ensuring load balance. This is a known NP-hard problem. We employ a greedy %
longest-processing-time-first scheduling (LPT) 
approach%
~\cite{graham1969bounds}. In this context, the ``processing time'' refers to the weight of each sub-metatree. We prioritize the ``heaviest'' sub-metatrees and assign them to the least loaded partitions. 
Sub-metatrees are sorted in descending order by their weights %
(line 10) %
and assigned to partitions with the smallest current load (i.e., total weight sum of sub-metatrees assigned to the partition) (lines 13-17).   
In Figure~\ref{fig:meta_partitioning_workflow} (Step 3), %
sub-metatrees $S_1$ and $S_2$ are assigned to Partition 1, and sub-metatree $S_3$ is assigned to Partition 2. %

\vspace{1mm}
\noindent\textit{Step 4: Deduplication.} %
Since the metatree can contain duplicated relations due to possible cycles in the metagraph, duplicates must be removed within each partition (lines 18-21). 
Each final HetG partition thus contains the complete mono-relation subgraph for each unique relation, including all nodes of the two node types connected by that relation and all edges of that edge type, %
as well as node features for its unique node types. %
In Figure~\ref{fig:meta_partitioning_workflow} (Step 4), Partition 2 has two instances of the ``Paper-cites-paper'' relation, and one of them is removed through deduplication. The final partitions are Partition 1, with five unique relations, 1.9 million nodes, and 30 million edges, and Partition 2, with three unique relations, 1.9 million nodes, and 25 million edges. 

\vspace{1mm}
\noindent\textbf{Complexity.} Our meta-partitioning operates on the metagraph rather than directly on nodes and edges of the HetG. The metagraph, which contains only a few vertices and links, is significantly smaller than the HetG.
The complexity of the BFS algorithm is at most $O(|R|)$ when it traverses all links. Since there are at most $|A|$ sub-metatrees (when all node types are connected to the target node type), the time complexity of sorting and deduplication is $O(|A| \log |A|)$. Therefore, the overall time complexity of the meta-partitioning algorithm is $O(|A| \log |A|) + O(|R|)$, which is much lower than graph partitioning algorithms in existing GNN training systems %
(at least $O(|V|)$, where $|V|$ is the number of nodes in the HetG~\cite{bulucc2016recent}). %

\noindent\textbf{Discussions.}
All existing benchmark HetG datasets~\cite{hu2020open, hu2021ogblsc, han2022openhgnn, donor} contain mono-relation subgraphs that fit in a single machine with 200 GB host memory, making our designs ideal for these cases. If a mono-relation subgraph is too large to fit in one machine, we can use a heuristic to find a balanced vertex-cut \cite{petroni2015hdrf}, referring to the problem (\ref{eq:graph_partition}). Then, we can still adapt the RAF paradigm (\S\ref{sec:computation_model}) to work across these partitions: each partition performs local aggregations on its portion of the subgraph, exchanges partial aggregation results for the cut vertices, and combines the local and received partial results to complete the relation-specific aggregation. After processing all relations spanning multiple machines, we proceed with the cross-relation aggregation as in the original RAF paradigm. 

When there are more partitions (machines) than sub-metatrees, we can duplicate sub-metatrees to use the %
$p$ machines. Replicated partitions will split the target nodes and train the same part of the HGNN model with data parallelism.

\section{%
GPU Feature Cache}%
\label{sec:gpu_cache}
We present the miss-penalty-aware cache size allocation and how we extend caching to include mutable learnable features and optimizer states and maintain cache consistency.

\begin{figure}[t]
    \centering
    \begin{subfigure}{0.49\columnwidth}
        \centering
        \includegraphics[width=\columnwidth]{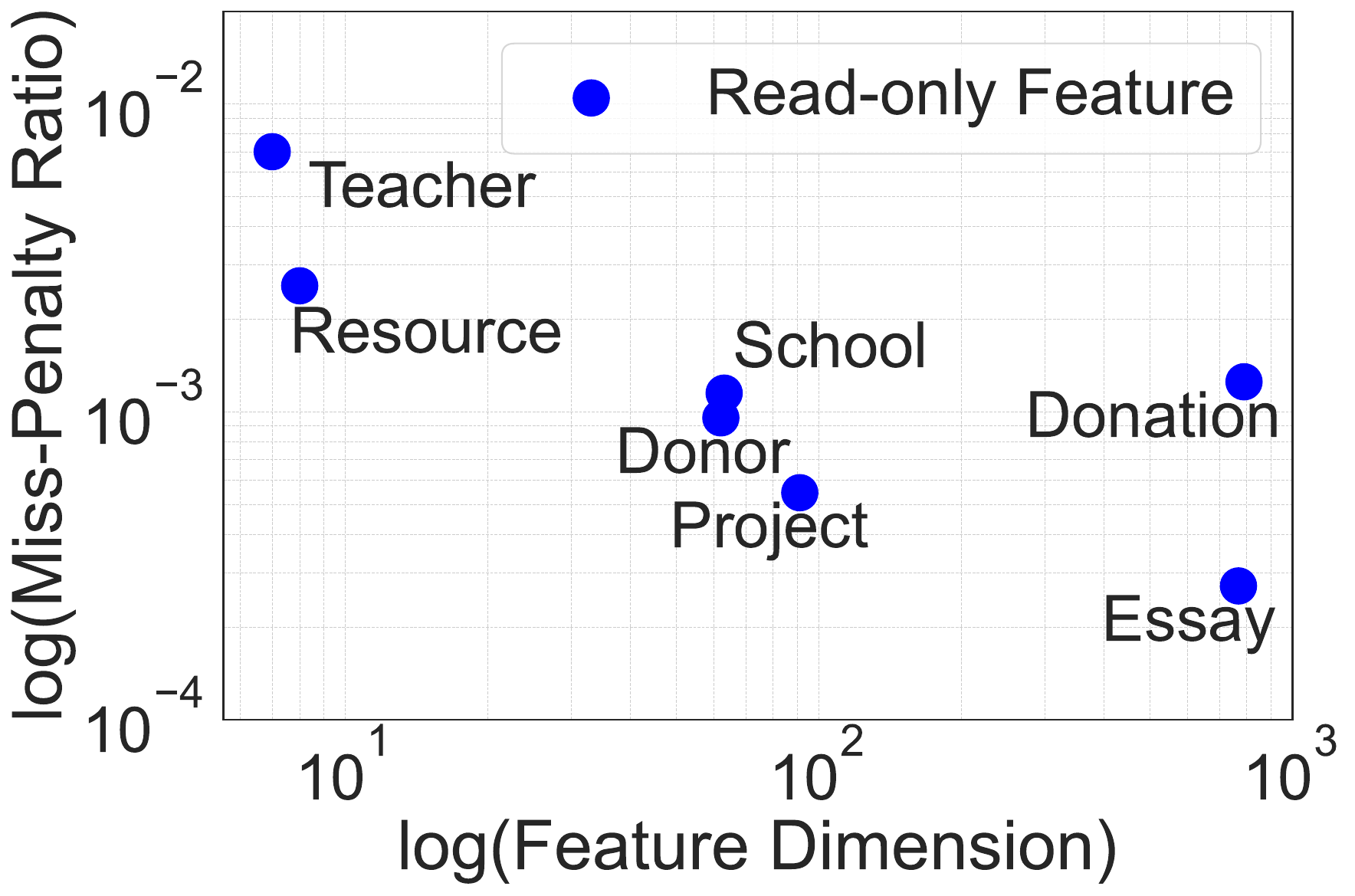}
        \vspace{-5mm}
        \caption{Donor dataset}
        \vspace{-3mm}
        \label{fig:miss_penalty_donor}
    \end{subfigure}
    \centering
    \begin{subfigure}{0.49\columnwidth}
        \centering
        \includegraphics[width=\columnwidth]{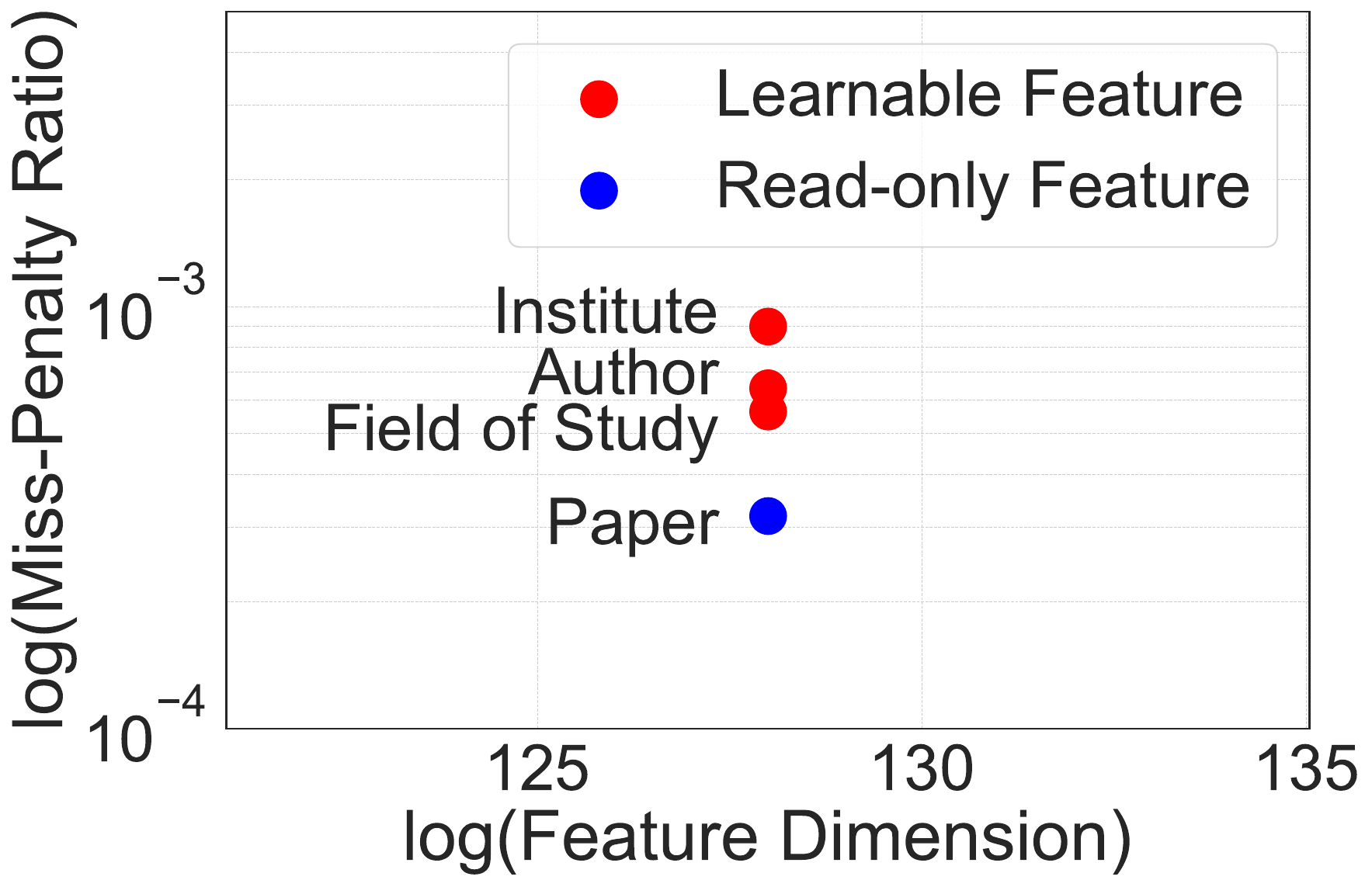}
        \vspace{-5mm}
        \caption{ogbn-mag dataset}
        \vspace{-3mm}
        \label{fig:miss_penalty_ogbn_mag}
    \end{subfigure}
    \hfill
    \caption{Miss-penalty ratio for different node types in Donor~\cite{donor} and ogbn-mag~\cite{hu2020open} datasets.} %
    \label{fig:miss_penalty_ratio}
\end{figure}

\vspace{1mm}
\noindent\textbf{%
Cache Size Allocation.} %
We design a cache size allocation strategy based jointly on the hotness of nodes and the \textit{miss-penalty ratio} of each node type. To determine node hotness, we adopt the pre-sampling method~\cite{yang2022gnnlab}, sampling nodes for two epochs before training and recording the number of times each node is sampled. As observed in prior works~\cite{lin2020pagraph, yang2022gnnlab}, we also find a skewed distribution of access frequency of different nodes in each node type. 
We define the miss-penalty ratio, denoted as $o_{a}$, which quantifies the time penalty per unit cache size incurred when the feature of a node of type $a$ is not in the cache. 
We profile the miss-penalty ratio for each node type before training. For read-only node features, we measure the time required to transfer features from DRAM to GPU memory. For learnable features, which also require caching of their optimizer states (e.g., moment and variance in the Adam optimizer~\cite{kingma2014adam}) due to the large overhead of updating them, we profile both the read and write times for these features and optimizer states, and then divide them by their cache size to calculate the ratio.  

Figure~\ref{fig:miss_penalty_donor} shows the miss-penalty ratios (microseconds per byte) for different node types with varying feature dimensions in the Donor dataset~\cite{donor}. It demonstrates that miss-penalty ratios vary significantly, with smaller feature dimensions generally corresponding to larger miss-penalty ratios due to fixed overhead per transfer %
(e.g., setting up the PCIe transaction)~\cite{sun2023legion}. Figure~\ref{fig:miss_penalty_ogbn_mag} illustrates the impact of learnable features, where we set the same feature dimension (128) for both learnable features and read-only node features. Learnable features exhibit larger miss-penalty ratios compared to read-only node features with the same dimension size, as they require additional write operations to DRAM. Typically, a higher miss-penalty ratio for a node type indicates greater potential benefit from caching that node type (due to a higher cost associated with cache misses), suggesting that we should cache more nodes of that type to improve cost-effectiveness.

To avoid the overhead of calculating scores (e.g., visited count weighted by miss-penalty) for every node and to prevent any single node type from dominating the cache, we propose a hierarchical caching strategy: first determine the cache size for each node type and then fill the allocated cache according to the visited counts of nodes in each node type (i.e., nodes with the highest visited counts are prioritized for caching until the cache is full). %
We allocate the cache size to each node type $a$ on each partition in proportion to the product of the total visit count $count_a$ of nodes of type $a$ (obtained from pre-sampling) and the miss-penalty ratio $o_a$: given the total cache size (e.g., 4GB per GPU as in our experiments), the percentage of the cache size allocated to node type $a$ is $\frac{count_a \times o_a}  {\sum_{a'\in A'}count_{a'} \times o_{a'}}$, where $A'$ is the set of all node types in the HetG partition.

\begin{figure*}[t]
    \centering
    \begin{subfigure}{0.32\textwidth}
        \centering
        \includegraphics[width=\columnwidth]{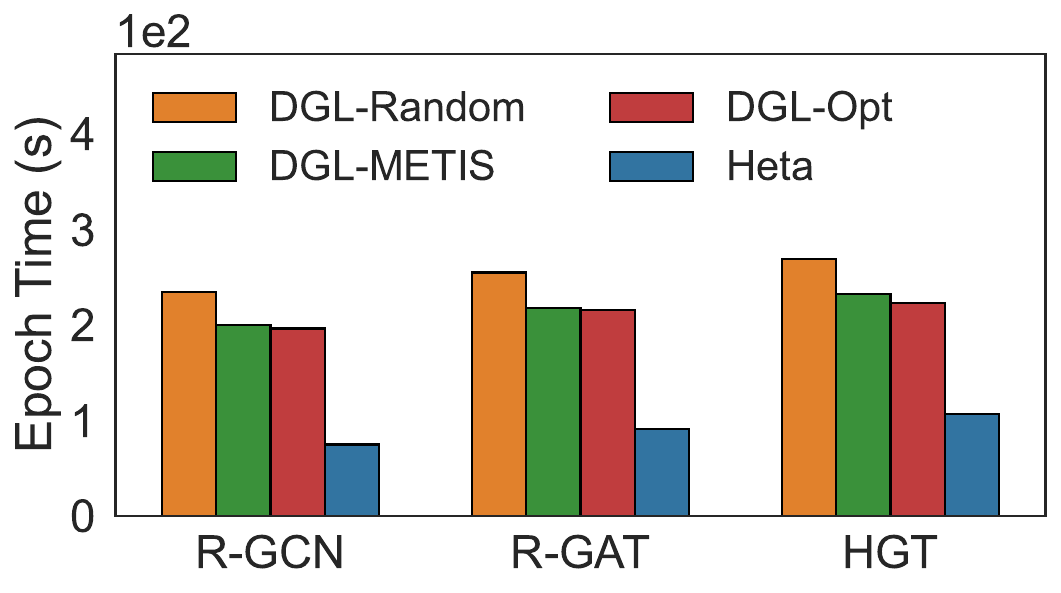}
        \vspace{-6mm}
        \caption{ogbn-mag dataset}
        \vspace{-3mm}
        \label{fig:ogbn_mag_overall_performance}
    \end{subfigure}
    \centering
    \begin{subfigure}{0.32\textwidth}
        \centering
        \includegraphics[width=\columnwidth]{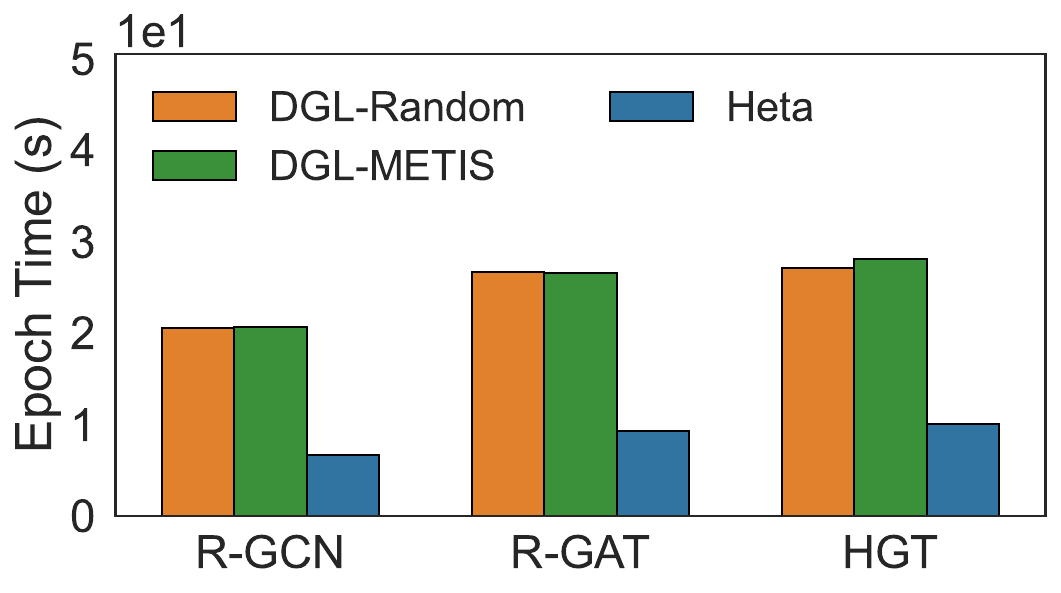}
        \vspace{-6mm}
        \caption{Freebase dataset}
        \vspace{-3mm}
        \label{fig:freebase_overall_performance}
    \end{subfigure}
    \centering
    \begin{subfigure}{0.32\textwidth}
        \centering
        \includegraphics[width=\columnwidth]{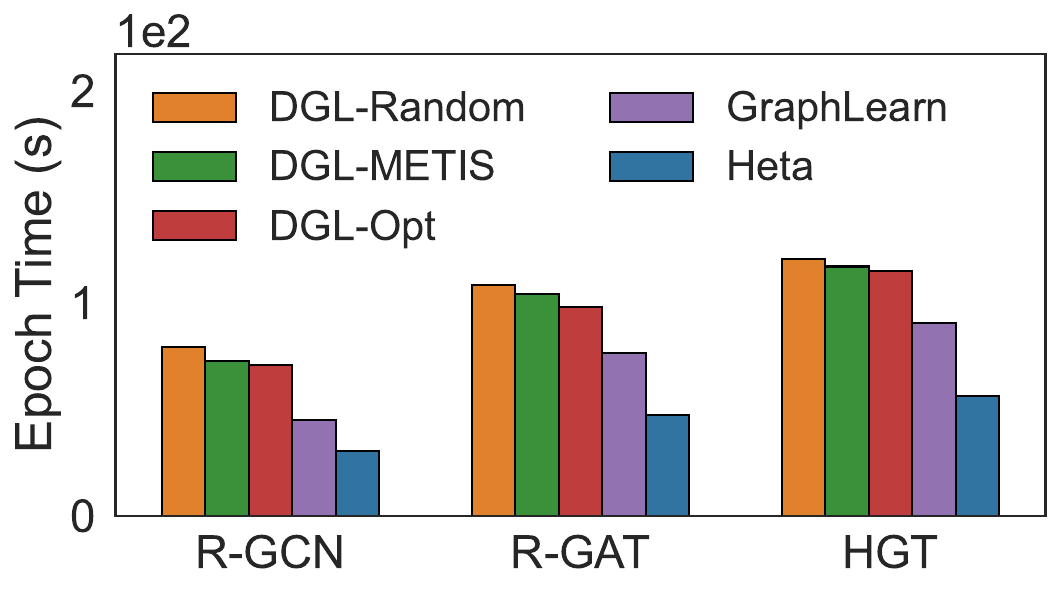}
        \vspace{-6mm}
        \caption{Donor dataset}
        \vspace{-3mm}
        \label{fig:donor_overall_performance}
    \end{subfigure}
    \caption{Overall performance on various HGNN models and medium-size datasets.} %
    \vspace{-3mm}
    \label{fig:end2end_epoch_time_small}
\end{figure*}

\vspace{1mm}
\noindent\textbf{Cache Consistency.}
Replicating %
learnable features in multiple GPU caches 
raises a consistency issue in the presence of writes. Any update to the features requires synchronization across all caches to ensure that all read operations retrieve the most recent version of the data.
To ensure cache consistency, %
we advocate a non-replicative cache split strategy to ensure that each learnable feature is
entirely stored in the GPU cache or the shared host memory, avoiding duplication and maintaining a single data version. 
In a multi-GPU setup within a single machine, the feature cache is partitioned across GPUs using a %
modular hashing scheme: %
the learnable feature of node $nid$ is cached in GPU rank $nid \mod num\_of\_gpus$. %
GPUs within the same machine can efficiently access each other's %
cache through CUDA peer-to-peer operations~\cite{cudap2p}. %

\section{Implementation}
\label{sec:impl}

We implement \sysname{} on DGL 1.1 and PyTorch 1.3.1, using about 2.8K lines of Python code. Minor modifications (around 300 lines) are made to DGL to enable %
GPU caching for features and optimizer states and to load graph partitions from meta-partitioning.
We use PyTorch's distributed package~\cite{li13pytorch} with Gloo~\cite{gloo} and NCCL~\cite{nccl} for gathering and all-reducing partial aggregations and gradients to implement the RAF paradigm. %
The meta-partitioning algorithm %
is implemented with NetworkX~\cite{networkx},  
saving necessary metadata for nodes/edges partitioning and storing the partitioned graph in DGL's format. The GPU cache is implemented in PyTorch with auxiliary data structures, including a 1D tensor for node ID recording and a 1D tensor to indicate node presence.
For multi-GPU machines, each machine is assigned a partition, and intra-machine workers perform data-parallel training by evenly splitting the target nodes. Each worker computes local relation aggregation on its target nodes and synchronizes gradients across workers in the same machine.  %

\sysname{} offers a user-friendly interface through three key APIs. 
The \texttt{Partition} function divides a graph into several partitions, with an optional argument for providing metapaths.
\texttt{FetchFeature} retrieves node/learnable features efficiently through GPU caching mechanisms.
The \texttt{HGNN} class allows users to define a HGNN model by specifying the relations, relation-specific aggregation functions ($\text{AGG}_r$), and a cross-relation aggregation function ($\text{AGG}_{\text{all}}$). %

\section{Evaluation}
\label{sec:evaluation}

{\small
\begin{table}[t]
\caption{Dataset information. \# Node T. and \# Edge T. represent the number of node types and edge types, respectively.} 
\vspace{-3.5mm}
\label{tab:dataset_characteristics}
\centering
\begin{tabular}{@{}llllll@{}}
\toprule
Attribute & ogbn-mag & Freebase & Donor & IGB-HET & MAG240M \\
\midrule
\# Nodes & 1.9e6 & 1.2e7 & 9.7e6 & 2.6e7 & 2.4e8 \\
\# Node T. & 4 & 8 & 7 & 4 & 3 \\
\# Edges & 4.2e7 & 1.3e8 & 2.5e7 & 4.9e8 & 3.4e9 \\
\# Edge T. & 7 & 64 & 14 & 7 & 5 \\
\makecell[l]{\# Node T. \\\quad w/ Feat.} & 1 & 0 & 7 & 4 & 1 \\
Feat. dim & 128 & N/A & 7-789 & 1024 & 768 \\
\# Classes & 349 & 8 & 2 & 2983 & 153 \\
Storage (GB) & 0.86 & 1.2 & 22 & 104 & 202 \\
\bottomrule
\end{tabular}
\end{table}
}

\subsection{Methodology}

\noindent\textbf{Testbed.} We conduct all experiments on Amazon EC2 g4dn.metal instances, each equipped with eight NVIDIA 16 GB T4 GPUs, 96 vCPU cores, and 384 GB DRAM. All instances are interconnected with a 100 Gbps network. By default, two instances are used in each experiment. We also use an Amazon EC2 X1.32xlarge instance with 2 TB DRAM for graph partitioning.

\vspace{1mm}
\noindent\textbf{Datasets.} We use %
heterogeneous graph datasets with the information given in Table~\ref{tab:dataset_characteristics}. 
The ogbn-mag dataset~\cite{hu2020open} comprises a heterogeneous bibliographic graph (Figure~\ref{fig:metagraph}).  %
The Freebase dataset~\cite{han2022openhgnn} is a knowledge graph extracted from Freebase~\cite{freebase:datadumps}. %
The Donor dataset~\cite{donor}, constructed using the method in~\cite{cvitkovic2020supervised}, comprises project applications submitted to DonorsChoose.org. %
The %
IGB-HET dataset is a heterogeneous citation network from the Illinois Graph Benchmark with many labeled nodes~\cite{khatua2023igb}.  %
The MAG240M~\cite{hu2021ogblsc} dataset originated from the Microsoft Academic Graph~\cite{wang2020microsoft} is the largest among the five. The numbers of classes for node classification tasks (\# Classes) are shown in the table.  %

\vspace{1mm}
\noindent\textbf{HGNNs.} We train three representative HGNN models: 
R-GCN~\cite{schlichtkrull2018modeling}, R-GAT~\cite{busbridge2019relational}, and HGT~\cite{hu2020heterogeneous}. We use the same model hyper-parameters as OGB leaderboards~\cite{hu2020open}, e.g., 2 layers and 64 hidden neurons per layer.
We set the minibatch size to 1024 by default and sample two-hop neighbors with fanout %
\{25, 20\} in all experiments. For IGB-HET, we set the minibatch size to 4096 since it has more training nodes. We allocate a 4GB cache size per GPU in \sysname{}. %

\vspace{1mm}
\noindent\textbf{Baselines.} We compare %
\sysname{} with the following baselines: (1) Distributed version of DGL~\cite{zheng2022distributed}, including: (i) {\em DGL-Random}, which employs random graph partitioning; (ii) {\em DGL-METIS}, employing METIS for graph partitioning; (iii) {\em DGL-Opt}, optimized DGL-METIS with the GPU cache %
that only caches node features (caching non-replicative learnable features offers little benefit to DGL, as remote workers still need network communication to fetch them). DGL-Opt is not applicable to the Freebase dataset, which has no node features. (2) GraphLearn (PyTorch version)~\cite{graphlearn}, which extends PyG~\cite{fey2019fast} to multi-GPU and distributed settings.
\footnote{PyG itself does not support multi-GPU nor distributed training on partitioned graphs, and thus, we do not directly compare with it.} 
GraphLearn does not support learnable features; therefore, we only run it on Donor and IGB-HET datasets in which all nodes have features. %
GraphLearn supports caching node features for hot nodes. We apply the same cache size per GPU (4GB) and the same cache size allocation method (\S\ref{sec:gpu_cache}) in \sysname{} to DGL-Opt and GraphLearn.

\begin{figure}[t]
    \centering
    \begin{subfigure}{0.49\columnwidth}
        \centering
        \includegraphics[width=\columnwidth]{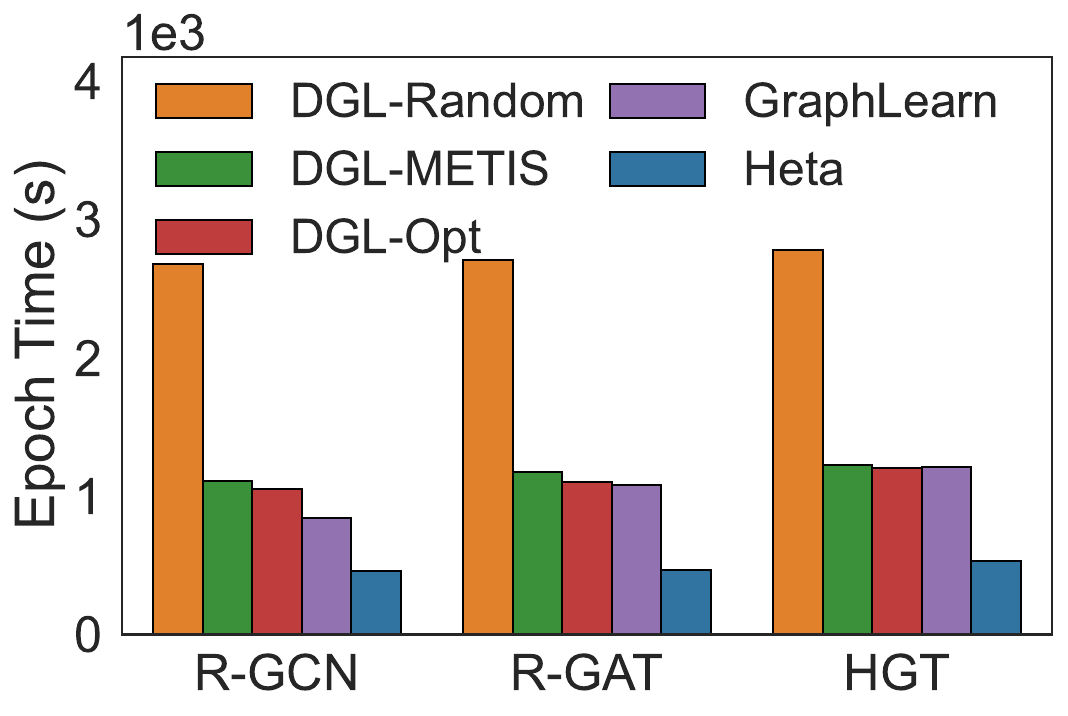}
        \vspace{-4mm}
        \caption{IGB-HET dataset}
        \vspace{-3mm}
        \label{fig:igb_het_overall_performance}
    \end{subfigure}
    \centering
    \begin{subfigure}{0.49\columnwidth}
        \centering
        \includegraphics[width=\columnwidth]{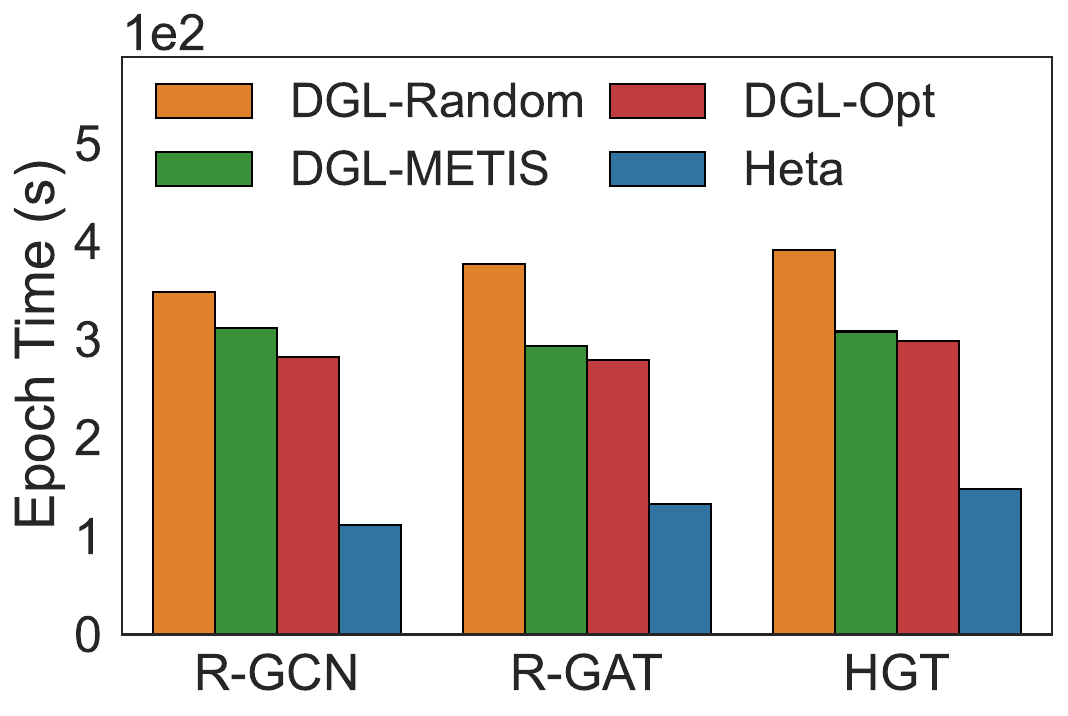}
        \vspace{-5mm}
        \caption{MAG240M dataset}
        \vspace{-3mm}
        \label{fig:mag240m_overall_performance}
    \end{subfigure}
    \hfill
    \caption{Overall performance on large datasets.} %
    \label{fig:end2end_epoch_time}
\end{figure}

\subsection{Overall Performance}

We first evaluate %
the epoch time of distributed training of \sysname{} and the baselines, i.e., the training time for one complete pass of the entire training dataset. %
As shown in %
Figure~\ref{fig:end2end_epoch_time_small} and Figure~\ref{fig:end2end_epoch_time},
\sysname{} significantly reduces the epoch time, %
achieving speedups ranging from 1.9$\times$ to 5.8$\times$ over DGL and 1.5$\times$ to 2.3$\times$ over GraphLearn.
GraphLearn performs the best among the baseline systems on the Donor and IGB-Het datasets, attributed to its asynchronous graph sampling and feature cache. %
\sysname{} outperforms all the baselines due to its RAF computation paradigm and meta-partitioning, effectively reducing cross-partition communication time. %
Besides, \sysname{} achieves %
an average speedup of 2.8$\times$, 2.6$\times$, and 2.4$\times$ compared to DGL and GraphLearn on R-GCN, R-GAT, and HGT, respectively. %
R-GAT and HGT are more computation-bound with the attention mechanism, while their communication is less intensive compared to R-GCN; a higher ratio of computation over other training stages leaves a smaller improvement space for \sysname{} because the communication bottleneck is less severe. %

\begin{figure}[t]
    \begin{subfigure}{\columnwidth}
        \centering
        \includegraphics[width=0.95\columnwidth]{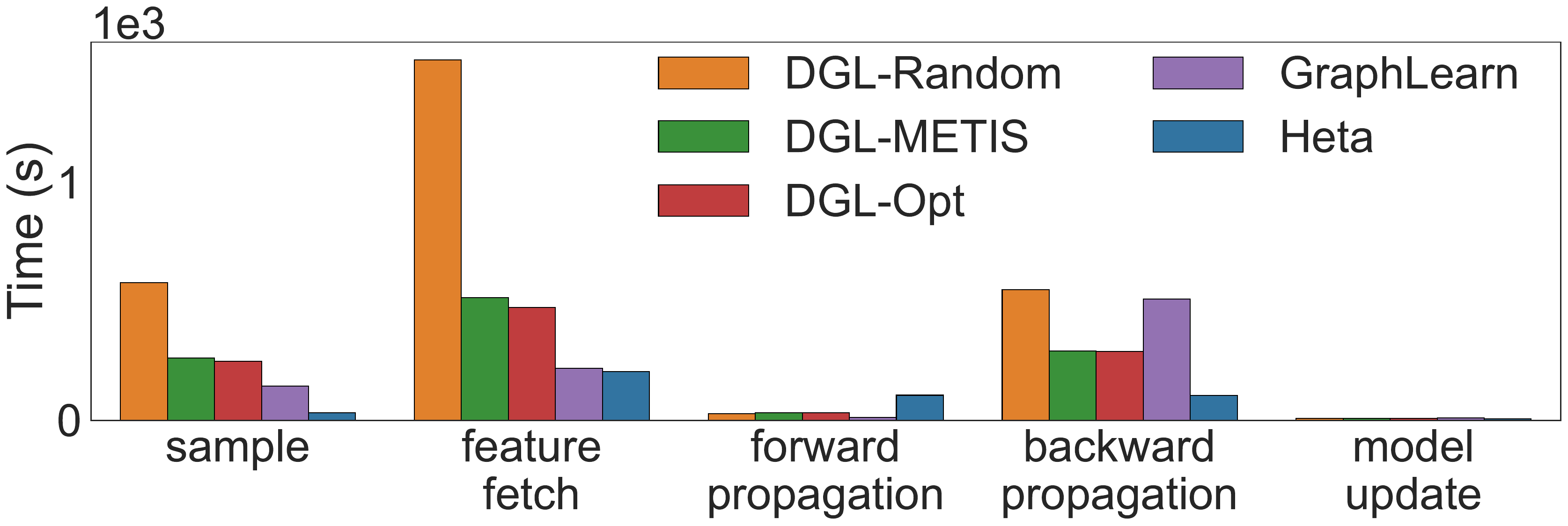}
        \vspace{-2mm}
        \caption{IGB-HET dataset}
        \label{fig:igb_het_performance_breakdown}
    \end{subfigure}
    \hfill
    \begin{subfigure}{\columnwidth}
        \centering
        \includegraphics[width=0.95\columnwidth]{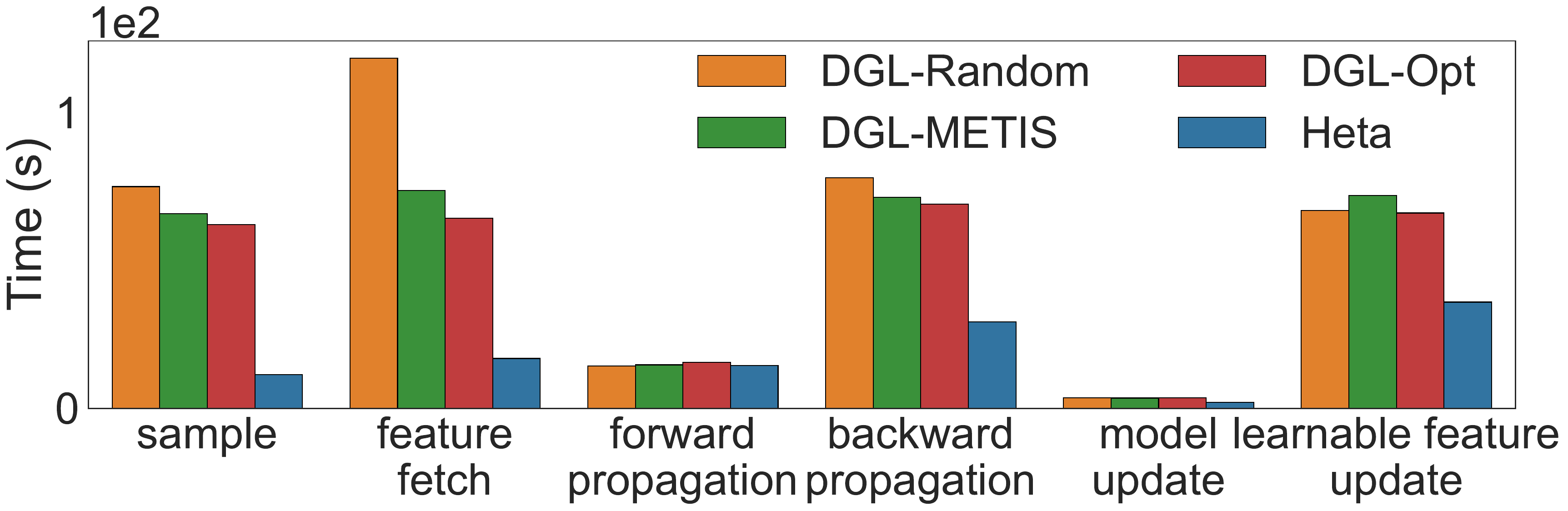}
        \vspace{-2mm}
        \caption{MAG240M dataset}
        \vspace{-3mm}
        \label{fig:mag240m_performance_breakdown}
    \end{subfigure}
    \caption{Time breakdown of training stages of R-GCN.} 
    \label{fig:breakdown_rgcn}
\end{figure}

\subsection{Training Time Breakdown}
We compare the time required for each training stage when training the R-GCN model. Figure~\ref{fig:breakdown_rgcn} presents the results on %
IGB-HET and MAG240M. Similar results were obtained on other models and datasets and omitted here due to space limitations. \sysname{} significantly reduces the time spent on sampling, feature fetching, and learnable feature and model updating. This %
is achieved by eliminating the need for cross-machine communication in these stages %
and executing them %
exclusively within the local partition, thanks to meta-partitioning. %
Furthermore, feature fetching and learnable feature update processes are optimized through the GPU cache. %
The forward computation time slightly increases because %
cross-partition communication is needed to aggregate relation-wise partial embeddings under the RAF paradigm. %
The backpropagation time is significantly reduced because %
\sysname{} does not require cross-machine gradient synchronization. In the case of multi-GPU machines, gradient synchronization among workers in the same machine is fast. The model update time is shortened %
since workers on each machine only hold a portion of the model with cross-machine model parallelism.

{\small
\begin{table}[t]
\centering
\caption{Partitioning Performance Comparison on MAG240M and IGB-HET datasets.} 
\vspace{-2mm}
\begin{tabular}{lcc|cc}
\toprule
\multirow{3}{*}{\textbf{Method}} & \multicolumn{2}{c}{\textbf{MAG240M}} & \multicolumn{2}{c}{\textbf{IGB-HET}} \\
\cmidrule(lr){2-3} \cmidrule(lr){4-5}
& \textbf{Time} & \textbf{Peak } & \textbf{Time} & \textbf{Peak} \\
& & \textbf{Memory} & & \textbf{Memory} \\
\midrule
Random & 3.9 hr & 917.3 GB & 1131 s & 253.0 GB \\
METIS & 7.2 hr & 847.7 GB & 3014 s & 261.9 GB \\
GraphLearn & - & - & 611 s & 175.4 GB \\
Meta-partitioning & 20.6 min & 406.3 GB & 549 s & 132.8 GB \\
\bottomrule
\end{tabular}
\label{tab:partition_comparison}
\end{table}
}

\subsection{Meta-Partitioning Efficiency}
Table~\ref{tab:partition_comparison} compares \sysname{}'s  %
meta-partitioning with random and METIS partitioning of DGL on the MAG240M and IGB-HET dataset (partitioning into 2 parts). %
Both random and METIS demand %
hours of partitioning time, %
mainly due to %
splitting the original HetG and shuffling nodes/edges to ensure contiguous ID ranges~\cite{zheng2022distributed}. In contrast, our meta-partitioning %
can be done in just 20.6 minutes. GraphLearn assumes all nodes have features for partitioning, making it unsuitable for MAG240M. When partitioning IGB-HET into two parts, GraphLearn's random partitioning takes 610 seconds, while random and METIS partitioning of DGL take 1131 and 3014 seconds, respectively. \sysname{} is the fastest, completing the task in 549 seconds: most of the time is spent saving partitioned graphs to disk, while building/splitting/assigning metatrees takes less than 1 second. Further, meta-partitioning reduces peak memory usage by eliminating extensive data reshuffling. 

\subsection{GPU Cache}

\begin{figure}[t]

    \begin{minipage}{0.48\columnwidth}
        \centering
        \vspace{-2mm}
        \includegraphics[width=\textwidth]{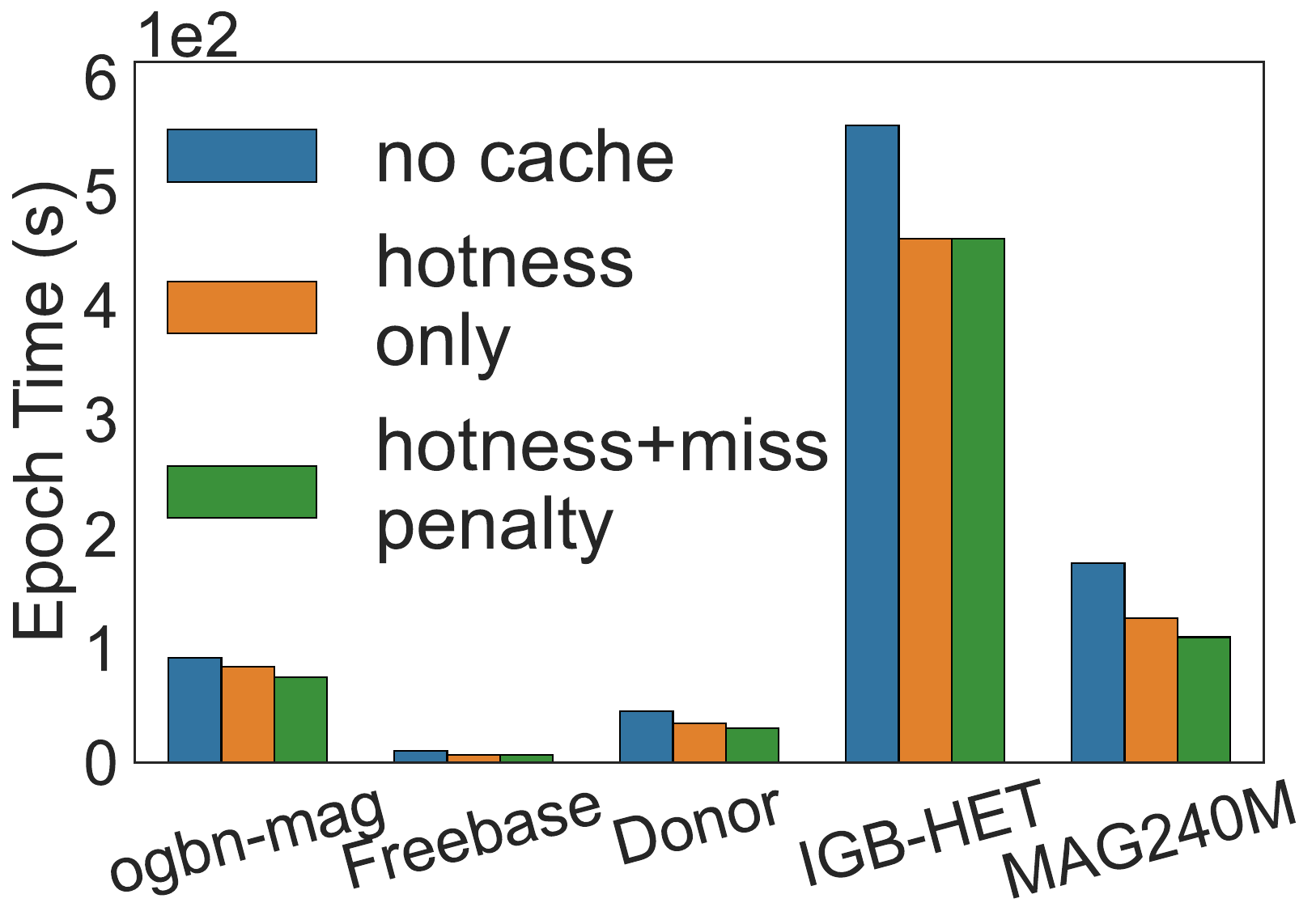}
        \vspace{-5mm}
        \caption{%
        Comparison of GPU cache methods.} 
        \label{fig:gpu_cache_ablation}
    \end{minipage}
    \hfill
    \begin{minipage}{0.48\columnwidth}
        \centering
        \vspace{-2mm}
        \includegraphics[width=\textwidth]{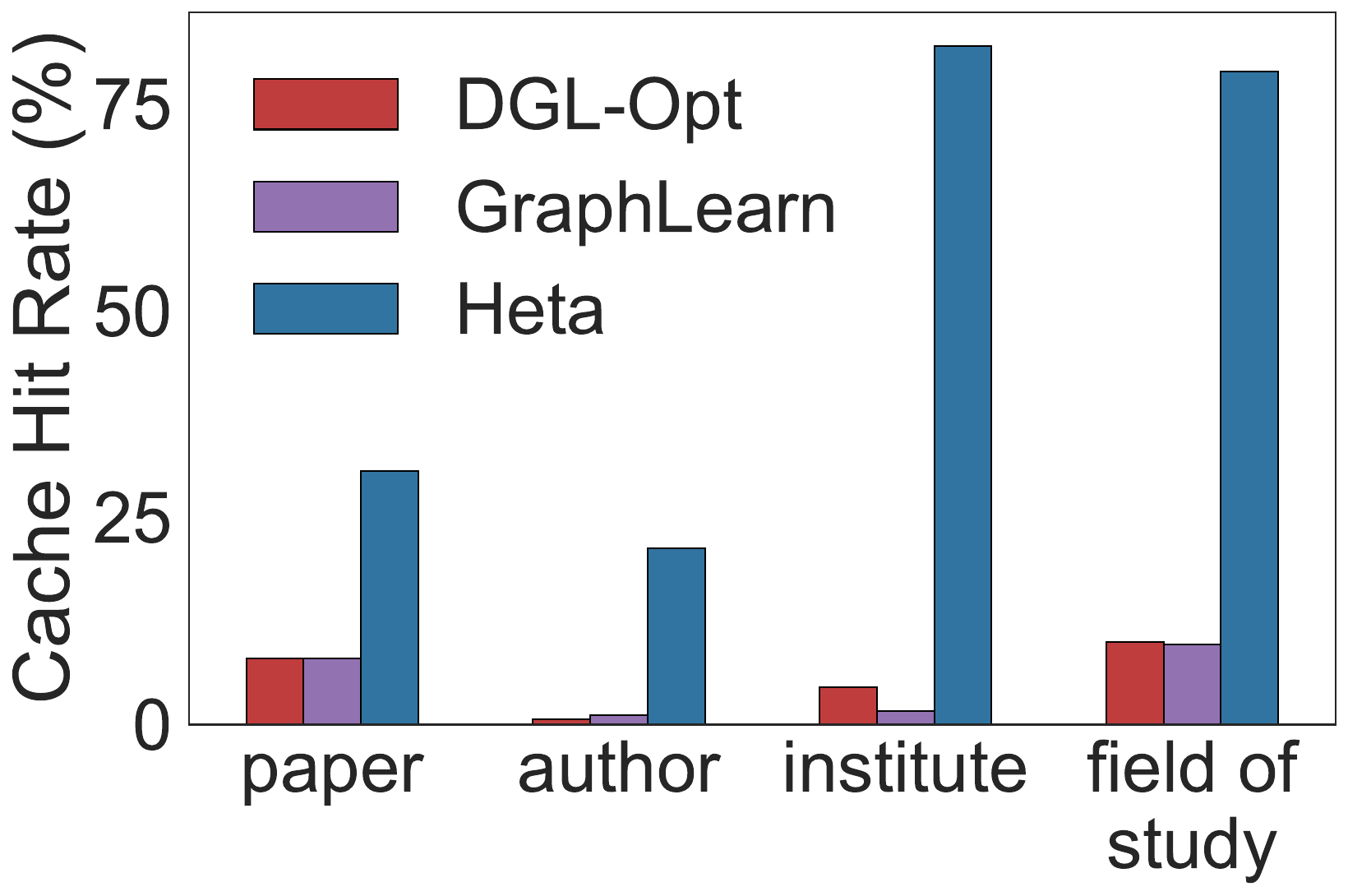}
        \vspace{-5mm}
        \caption{%
        Cache hit rate comparison on IGB-HET.}
        \label{fig:gpu_cache_hit_rate}
    \end{minipage}
\end{figure}

Figure~\ref{fig:gpu_cache_ablation} shows the epoch time in training R-GCN on various datasets. %
The `hotness only' design allocates cache size considering only the hotness of nodes, %
and `hotness+miss-penalty' is \sysname{}'s cache design considering both the hotness of nodes and the miss-penalty ratio. %
\sysname{}'s cache design achieves up to 1.6$\times$ training speedup than no-cache and up to 15\% faster than hotness-only caching. A more significant advantage is shown on MAG240M and Donor (15\% and 13\% compared to hotness only, respectively) when the miss-penalty is considered, due to their varied feature dimensions and learnable features (for MAG240M). 
The least benefit is shown on IGB-HET due to its uniform feature dimension. Varying feature dimensions and whether features are learnable can significantly impact the miss-penalty ratios, which in turn affects cache performance by altering the efficiency of data retrieval (\S\ref{sec:gpu_cache}).  %

Figure~\ref{fig:gpu_cache_hit_rate} shows that %
\sysname{}'s cache hit rates for all node types are substantially higher than those of DGL-Opt and GraphLearn when training R-GAT on IGB-HET. Similar results are observed on other datasets as well. This %
is largely because our meta-partitioning leads to limited node types in each partition, and allows the GPU caches to store only the node types needed for local computation. In contrast, DGL and GraphLearn must cache all types of nodes since sampled neighbors are from the whole graph. %
For example, we have observed a %
1.6$\times$ training speedup with \sysname{} %
than no cache in Figure~\ref{fig:gpu_cache_ablation}, as compared to DGL-Opt's 1.1$\times$ speedup (with GPU cache) than DGL-METIS (without GPU cache) in Figure~\ref{fig:mag240m_overall_performance}. %
This is because DGL-Opt only achieves a cache hit rate of 11\% for the ``paper'' node type in MAG240M, %
while \sysname{} achieves a %
67\% cache hit rate for the same node type. Consequently, \sysname{} significantly reduces the overhead of feature fetching.

\begin{figure}[t]
    \begin{minipage}{0.47\columnwidth}
        \centering
        \includegraphics[width=\columnwidth]{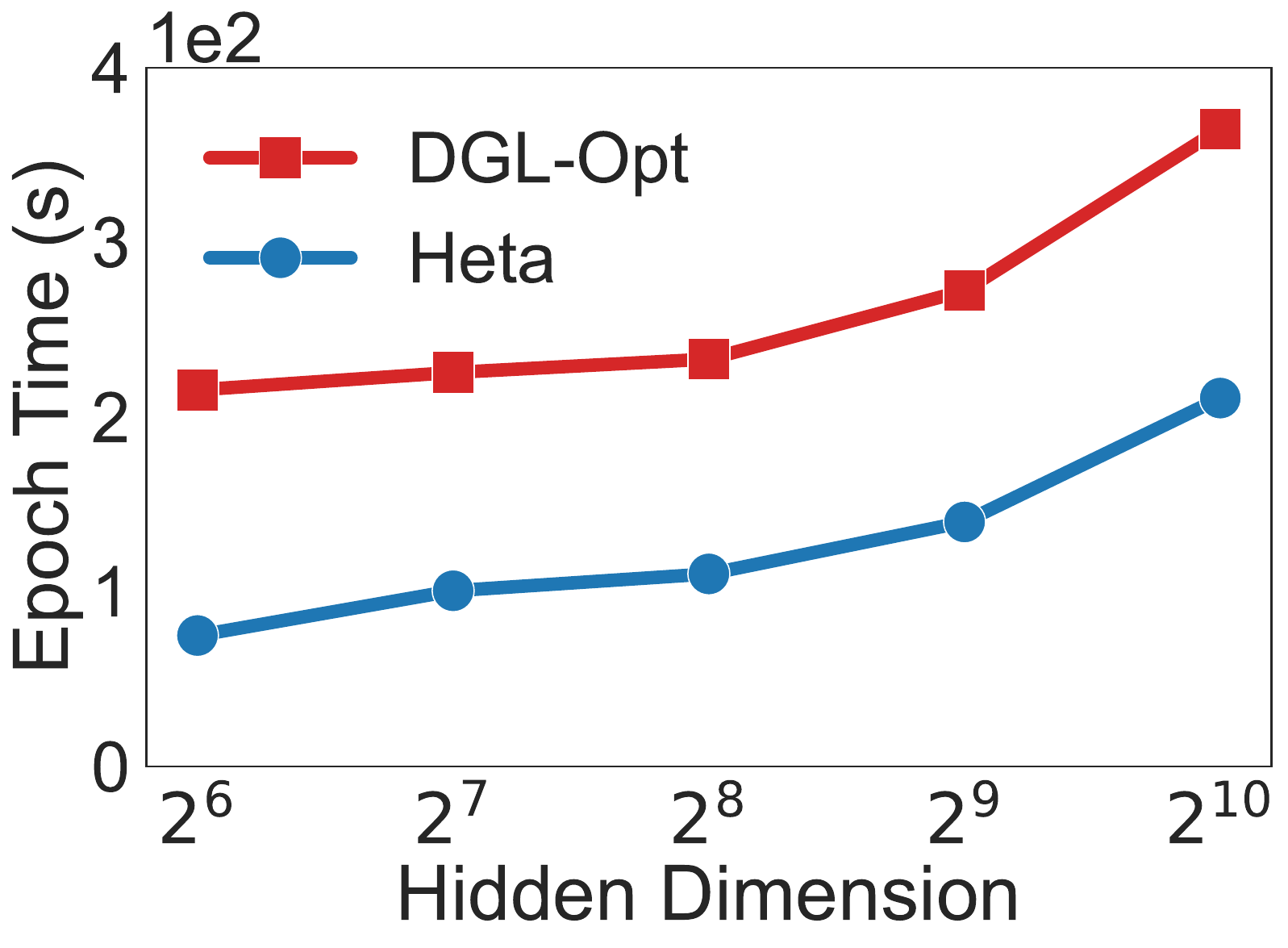}
        \vspace{-7mm}
        \caption{%
        Performance under different hidden dimension sizes.}
        \label{fig:hidden_dim_ablation}
    \end{minipage}
    \hfill
    \begin{minipage}{0.49\columnwidth}
        \centering
        \includegraphics[width=\columnwidth]{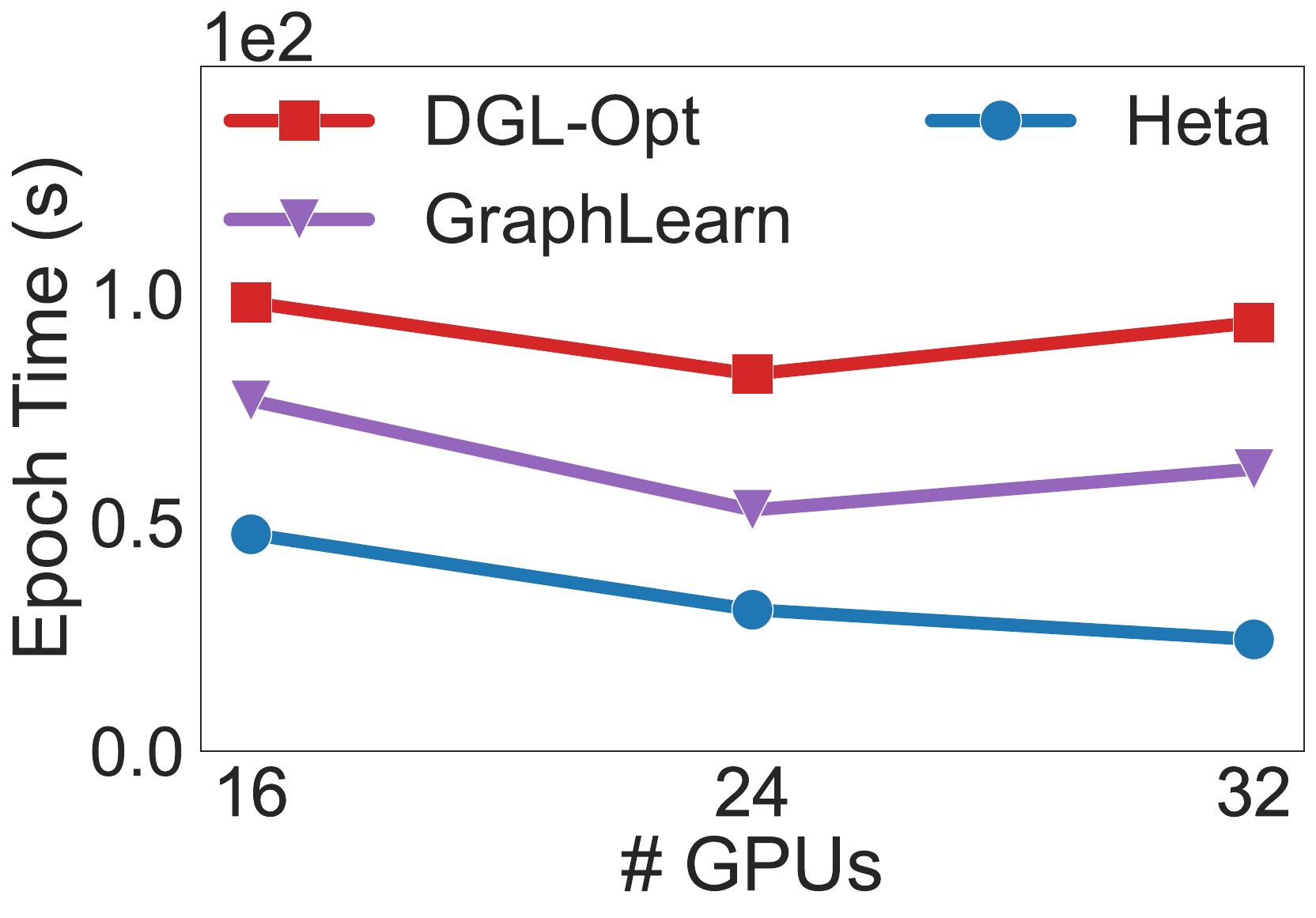}
        \vspace{-7mm}
        \caption{%
        Performance under different GPU numbers.
        }
        \label{fig:num_partition_ablation}
    \end{minipage}
\end{figure}

\subsection{Ablation Study}

\vspace{1mm}
\noindent\textbf{%
Hidden Dimension.}
In Figure~\ref{fig:hidden_dim_ablation}, we increase the hidden dimension from 64 to 1024 in the R-GCN model and evaluate the training epoch time %
on the ogbn-mag dataset. %
As the hidden dimension increases, network communication in \sysname{} increases due to collecting more partial aggregations, 
leading to increased epoch time. %
Nonetheless, \sysname{} still achieves a 1.7$\times$ speedup than DGL-Opt when the hidden dimension is 1024. It is worth noting that the hidden dimension in HGNN models is typically smaller than the feature dimension and does not reach such large values~\cite{wang2022survey}. %

\vspace{1mm}
\noindent\textbf{%
Scalability.} 
In Figure~\ref{fig:num_partition_ablation}, we evaluate the training epoch time of R-GAT on the Donor dataset with 16, 24, and 32 GPUs distributed across 2, 3, and 4 Amazon EC2 g4dn.metal instances, respectively.
We maintain the same global batch size of 1024 across the machines. \sysname{} exhibits excellent scalability as the number of GPUs increases. %
In contrast, the epoch time of DGL-Opt and GraphLearn increases when moving from 24 to 32 GPUs. %
Although the computation workload decreases due to fewer training nodes per GPU, the communication overhead in DGL-Opt and GraphLearn grows significantly as the graph data is spread among more machines, %
requiring more inter-machine data fetching. %
\sysname{} eliminates the issue through meta-partitioning, %
which constrains the boundary nodes to target nodes, making communication complexity constant.%

\begin{figure}[t]
    \centering
    \includegraphics[width=0.7\columnwidth]{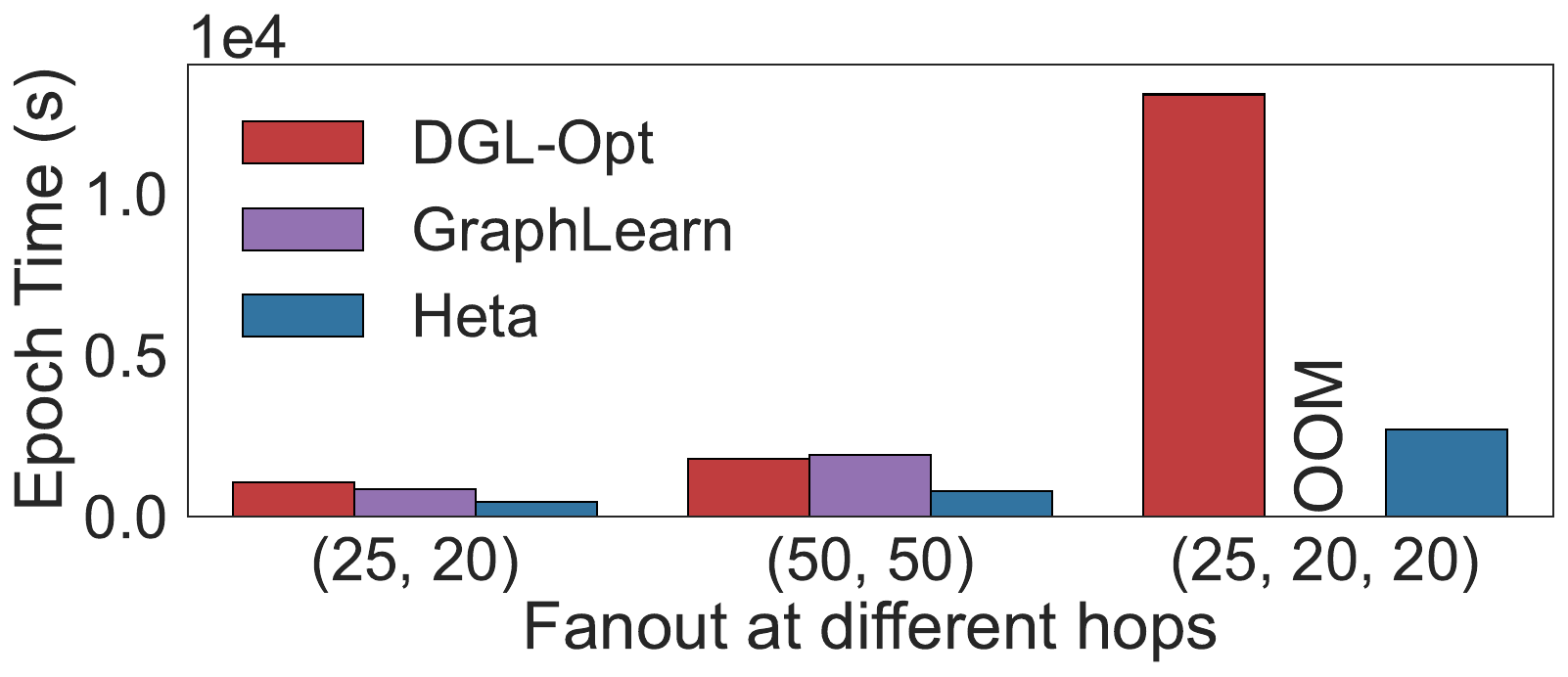}
    \vspace{-3mm}
    \caption{%
    Performance under different sampling fanouts and hops.} 
    \label{fig:sampling_ablation}
\end{figure}

\vspace{1mm}
\noindent\textbf{%
Sampling.}
In Figure~\ref{fig:sampling_ablation}, we experiment with different sampling fanouts and neighborhood hops when training R-GCN on the IGB-HET dataset. \sysname{} exhibits 2.3$\times$ to 4.9$\times$ speedups than DGL-Opt with larger fanout and more neighborhood hops. GraphLearn encountered out-of-memory issues with fanouts of \{25, 20, 20\} for three sampling hops. %
Cross-partition communication of \sysname{} remains constant due to meta-partitioning, %
while the baselines demand more network communication when sampling a larger neighborhood. %

\subsection{Model Accuracy}

\begin{figure}[t]
    \begin{subfigure}{0.49\columnwidth}
        \includegraphics[width=\columnwidth]{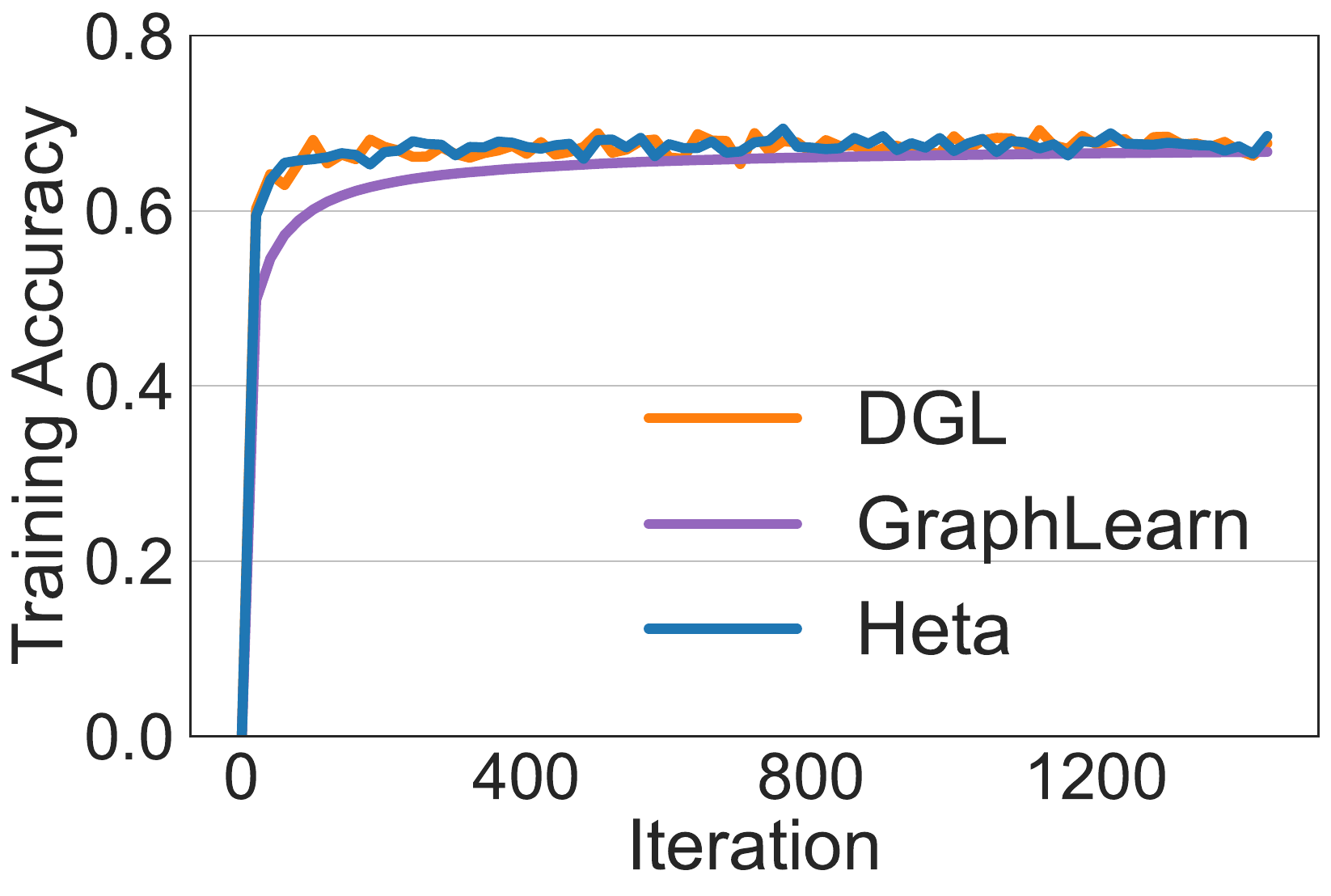}
        \vspace{-5mm}
        \caption{IGB-HET dataset}
        \vspace{-3mm}
        \label{fig:igb_het_accuracy}
    \end{subfigure}
    \hfill
    \begin{subfigure}{0.49\columnwidth}
        \includegraphics[width=\columnwidth]{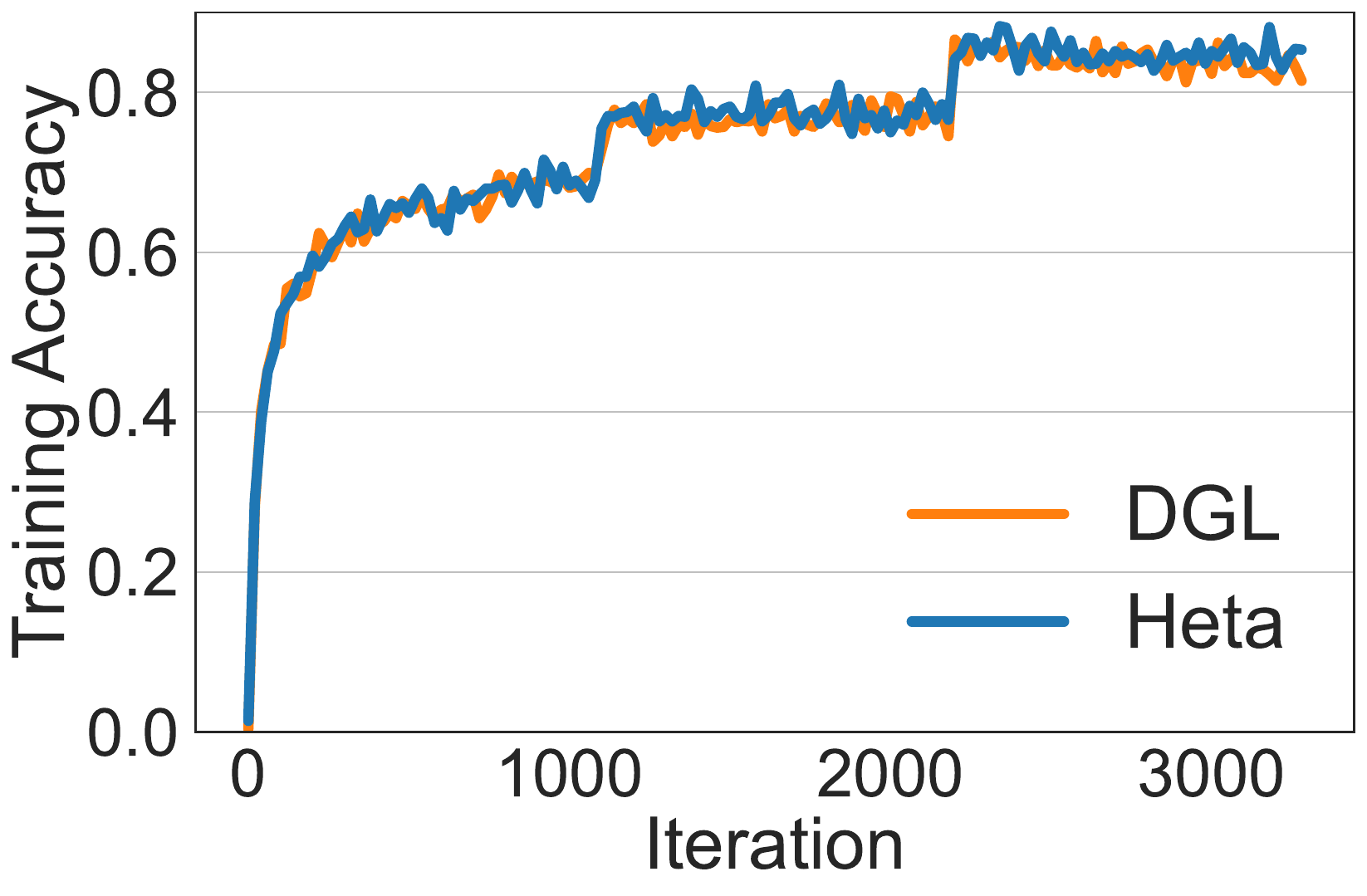}
        \vspace{-5mm}
        \caption{MAG240M dataset}
        \vspace{-3mm}
        \label{fig:mag240m_accuracy}
    \end{subfigure}
    \caption{\sysname{} achieves the same model accuracy as DGL.}
    \label{fig:accuracy}
\end{figure}

Figure~\ref{fig:igb_het_accuracy} depicts the training accuracy curves of \sysname{} and the baseline systems when training R-GAT on the IGB-HET dataset, and Figure~\ref{fig:mag240m_accuracy} presents those for training HGT on MAG240M. %
\sysname{} achieves the same training accuracy as DGL, while GraphLearn falls short of matching DGL's accuracy on the IGB-HET dataset. Regarding the test accuracy, both \sysname{} and DGL achieve a score of 0.66 on the MAG240M dataset, which is consistent with the leaderboard results~\cite{ogblscleaderboard}. %
All three systems attain the same accuracy of 0.68 on the IGB-HET dataset, which aligns with the reported accuracy in their original paper~\cite{khatua2023igb}.  These results align with our theoretical analysis in \S\ref{sec:computation_model} that \sysname{} does not alter the mathematical equivalence of the model.

\section{Related Work}

\vspace{1mm}
\noindent\textbf{Graph Partitioning in Graph Processing Systems.} Pregel~\cite{malewicz2010pregel} and GraphX~\cite{gonzalez2014graphx} adopt edge-cut graph partitioning. PowerGraph~\cite{gonzalez2012powergraph} adopts vertex-cut graph partitioning. %
PowerLyra~\cite{chen2019powerlyra} %
combines the benefits of vertex-cut and edge-cut partitioning for power-law graphs. %
They consider property graphs %
but ignore the property information during partitioning (i.e., treating a property graph as homogeneous). METIS~\cite{metis1998} is a multi-level, multi-constraint graph partitioning method. %
It is hard to achieve our meta-partitioning effects since it does not consider diverse nodes and edge types present in HetGs and the inherent structure of HGNNs. It also aims to minimize the number of edges cut rather than boundary nodes.  %

\vspace{1mm}
\noindent\textbf{GPU Feature Cache.}
PaGraph~\cite{lin2020pagraph} and GNNLab~\cite{yang2022gnnlab} advocate %
GPU cache for read-only features. %
BGL~\cite{liu2021bgl} adopts dynamic node caching using the FIFO policy. Legion~\cite{sun2023legion} proposes a unified GPU cache for both features and graph topology. GNNFlow~\cite{zhong2023gnnflow} advocates dynamic cache (e.g., LRU) for both node and edge features. None of them consider different cache miss penalties for different node types nor updatable learnable features.

\section{Conclusion}
\sysname{} is an efficient distributed framework %
for distributed HGNN training on heterogeneous graphs. Employing the RAF computation paradigm and meta-partitioning, \sysname{} confines the boundary nodes to the target node type and achieves lower communication complexity. \sysname{}'s miss-penalty-aware GPU feature cache further reduces communication. Our comprehensive evaluation demonstrates that \sysname{} achieves superior performance, up to 5.8$\times$ speedup in epoch time without loss of accuracy, compared to state-of-the-art systems DGL and GraphLearn. \sysname{}'s performance gains become more pronounced as the sampling neighborhood size increases. Also, \sysname{}'s meta-partitioning is more efficient in time and memory footprint than state-of-the-art graph partitioning methods such as METIS.

\bibliographystyle{ACM-Reference-Format}
\bibliography{citation.bib}


\begin{thebibliography}{55}


\ifx \showCODEN    \undefined \def \showCODEN     #1{\unskip}     \fi
\ifx \showDOI      \undefined \def \showDOI       #1{#1}\fi
\ifx \showISBNx    \undefined \def \showISBNx     #1{\unskip}     \fi
\ifx \showISBNxiii \undefined \def \showISBNxiii  #1{\unskip}     \fi
\ifx \showISSN     \undefined \def \showISSN      #1{\unskip}     \fi
\ifx \showLCCN     \undefined \def \showLCCN      #1{\unskip}     \fi
\ifx \shownote     \undefined \def \shownote      #1{#1}          \fi
\ifx \showarticletitle \undefined \def \showarticletitle #1{#1}   \fi
\ifx \showURL      \undefined \def \showURL       {\relax}        \fi
\providecommand\bibfield[2]{#2}
\providecommand\bibinfo[2]{#2}
\providecommand\natexlab[1]{#1}
\providecommand\showeprint[2][]{arXiv:#2}

\bibitem[Benchmark(2024)]%
        {ogblscleaderboard}
\bibfield{author}{\bibinfo{person}{Open~Graph Benchmark}.} \bibinfo{year}{2024}\natexlab{}.
\newblock \bibinfo{title}{OGB-LSC Leaderboards}.
\newblock \bibinfo{howpublished}{\url{https://ogb.stanford.edu/docs/lsc/leaderboards/}}.
\newblock


\bibitem[Bulu{\c{c}} et~al\mbox{.}(2016)]%
        {bulucc2016recent}
\bibfield{author}{\bibinfo{person}{Ayd{\i}n Bulu{\c{c}}}, \bibinfo{person}{Henning Meyerhenke}, \bibinfo{person}{Ilya Safro}, \bibinfo{person}{Peter Sanders}, {and} \bibinfo{person}{Christian Schulz}.} \bibinfo{year}{2016}\natexlab{}.
\newblock \bibinfo{booktitle}{\emph{{Recent Advances in Graph Partitioning}}}.
\newblock \bibinfo{publisher}{Springer}.
\newblock


\bibitem[Busbridge et~al\mbox{.}(2019)]%
        {busbridge2019relational}
\bibfield{author}{\bibinfo{person}{Dan Busbridge}, \bibinfo{person}{Dane Sherburn}, \bibinfo{person}{Pietro Cavallo}, {and} \bibinfo{person}{Nils~Y Hammerla}.} \bibinfo{year}{2019}\natexlab{}.
\newblock \showarticletitle{{Relational Graph Attention Networks}}.
\newblock \bibinfo{journal}{\emph{arXiv preprint}} (\bibinfo{year}{2019}).
\newblock


\bibitem[Chen et~al\mbox{.}(2015)]%
        {chen2019powerlyra}
\bibfield{author}{\bibinfo{person}{Rong Chen}, \bibinfo{person}{Jiaxin Shi}, \bibinfo{person}{Yanzhe Chen}, {and} \bibinfo{person}{Haibo Chen}.} \bibinfo{year}{2015}\natexlab{}.
\newblock \showarticletitle{{PowerLyra: Differentiated Graph Computation and Partitioning on Skewed Graphs}}. In \bibinfo{booktitle}{\emph{Proceedings of the Tenth European Conference on Computer Systems}}.
\newblock


\bibitem[Cvitkovic(2020)]%
        {cvitkovic2020supervised}
\bibfield{author}{\bibinfo{person}{Milan Cvitkovic}.} \bibinfo{year}{2020}\natexlab{}.
\newblock \showarticletitle{{Supervised Learning on Relational Databases with Graph Neural Networks}}.
\newblock \bibinfo{journal}{\emph{arXiv preprint}} (\bibinfo{year}{2020}).
\newblock


\bibitem[Developers(2023)]%
        {networkx}
\bibfield{author}{\bibinfo{person}{NetworkX Developers}.} \bibinfo{year}{2023}\natexlab{}.
\newblock \bibinfo{title}{{NetworkX}}.
\newblock
\newblock
\newblock
\shownote{\url{https://networkx.org/}}.


\bibitem[Facebook(2023)]%
        {gloo}
\bibfield{author}{\bibinfo{person}{Facebook}.} \bibinfo{year}{2023}\natexlab{}.
\newblock \bibinfo{title}{{Gloo}}.
\newblock
\newblock
\newblock
\shownote{\url{https://github.com/facebookincubator/gloo}}.


\bibitem[Fey and Lenssen(2019)]%
        {fey2019fast}
\bibfield{author}{\bibinfo{person}{Matthias Fey} {and} \bibinfo{person}{Jan~Eric Lenssen}.} \bibinfo{year}{2019}\natexlab{}.
\newblock \showarticletitle{{Fast Graph Representation Learning with PyTorch Geometric}}. In \bibinfo{booktitle}{\emph{Proceedings of ICLR workshop on Representation Learning on Graphs and Manifolds}}.
\newblock


\bibitem[Fey et~al\mbox{.}(2021)]%
        {fey2021gnnautoscale}
\bibfield{author}{\bibinfo{person}{Matthias Fey}, \bibinfo{person}{Jan~E Lenssen}, \bibinfo{person}{Frank Weichert}, {and} \bibinfo{person}{Jure Leskovec}.} \bibinfo{year}{2021}\natexlab{}.
\newblock \showarticletitle{{GNNAutoScale: Scalable and Expressive Graph Neural Networks via Historical Embeddings}}. In \bibinfo{booktitle}{\emph{Proceedings of International Conference on Machine Learning}}.
\newblock


\bibitem[Gandhi and Iyer(2021)]%
        {gandhi2021p3}
\bibfield{author}{\bibinfo{person}{Swapnil Gandhi} {and} \bibinfo{person}{Anand~Padmanabha Iyer}.} \bibinfo{year}{2021}\natexlab{}.
\newblock \showarticletitle{{{$P^3$}: Distributed Deep Graph Learning at Scale}}. In \bibinfo{booktitle}{\emph{Proceedings of the 15th USENIX Symposium on Operating Systems Design and Implementation}}.
\newblock


\bibitem[Gao et~al\mbox{.}(2022)]%
        {gao2022hetinf}
\bibfield{author}{\bibinfo{person}{Liqun Gao}, \bibinfo{person}{Haiyang Wang}, \bibinfo{person}{Zhouran Zhang}, \bibinfo{person}{Hongwu Zhuang}, {and} \bibinfo{person}{Bin Zhou}.} \bibinfo{year}{2022}\natexlab{}.
\newblock \showarticletitle{{HetInf: Social Influence Prediction with Heterogeneous Graph Neural Network}}.
\newblock \bibinfo{journal}{\emph{Frontiers in Physics}} (\bibinfo{year}{2022}).
\newblock


\bibitem[Gonzalez et~al\mbox{.}(2012)]%
        {gonzalez2012powergraph}
\bibfield{author}{\bibinfo{person}{Joseph~E Gonzalez}, \bibinfo{person}{Yucheng Low}, \bibinfo{person}{Haijie Gu}, \bibinfo{person}{Danny Bickson}, {and} \bibinfo{person}{Carlos Guestrin}.} \bibinfo{year}{2012}\natexlab{}.
\newblock \showarticletitle{{PowerGraph: Distributed Graph-Parallel Computation on Natural Graphs}}. In \bibinfo{booktitle}{\emph{Proceedings of the 10th USENIX Symposium on Operating Systems Design and Implementation}}.
\newblock


\bibitem[Gonzalez et~al\mbox{.}(2014)]%
        {gonzalez2014graphx}
\bibfield{author}{\bibinfo{person}{Joseph~E Gonzalez}, \bibinfo{person}{Reynold~S Xin}, \bibinfo{person}{Ankur Dave}, \bibinfo{person}{Daniel Crankshaw}, \bibinfo{person}{Michael~J Franklin}, {and} \bibinfo{person}{Ion Stoica}.} \bibinfo{year}{2014}\natexlab{}.
\newblock \showarticletitle{{GraphX: Graph Processing in a Distributed Dataflow Framework}}. In \bibinfo{booktitle}{\emph{Proceedings of the 11th USENIX Symposium on Operating Systems Design and Implementation}}.
\newblock


\bibitem[Google(2024)]%
        {freebase:datadumps}
\bibfield{author}{\bibinfo{person}{Google}.} \bibinfo{year}{2024}\natexlab{}.
\newblock \bibinfo{title}{Freebase Data Dumps}.
\newblock \bibinfo{howpublished}{\url{https://developers.google.com/freebase/data}}.
\newblock


\bibitem[Graham(1969)]%
        {graham1969bounds}
\bibfield{author}{\bibinfo{person}{Ronald~L. Graham}.} \bibinfo{year}{1969}\natexlab{}.
\newblock \showarticletitle{{Bounds on Multiprocessing Timing Anomalies}}.
\newblock \bibinfo{journal}{\emph{SIAM journal on Applied Mathematics}} (\bibinfo{year}{1969}).
\newblock


\bibitem[Hamilton et~al\mbox{.}(2017)]%
        {hamilton2017inductive}
\bibfield{author}{\bibinfo{person}{Will Hamilton}, \bibinfo{person}{Zhitao Ying}, {and} \bibinfo{person}{Jure Leskovec}.} \bibinfo{year}{2017}\natexlab{}.
\newblock \showarticletitle{{Inductive Representation Learning on Large Graphs}}. In \bibinfo{booktitle}{\emph{Proceedings of Advances in Neural Information Processing Systems}}.
\newblock


\bibitem[Han et~al\mbox{.}(2022)]%
        {han2022openhgnn}
\bibfield{author}{\bibinfo{person}{Hui Han}, \bibinfo{person}{Tianyu Zhao}, \bibinfo{person}{Cheng Yang}, \bibinfo{person}{Hongyi Zhang}, \bibinfo{person}{Yaoqi Liu}, \bibinfo{person}{Xiao Wang}, {and} \bibinfo{person}{Chuan Shi}.} \bibinfo{year}{2022}\natexlab{}.
\newblock \showarticletitle{{OpenHGNN: An Open Source Toolkit for Heterogeneous Graph Neural Network}}. In \bibinfo{booktitle}{\emph{Proceedings of the 31st ACM International Conference on Information \& Knowledge Management}}.
\newblock


\bibitem[Hu et~al\mbox{.}(2019)]%
        {hu2019cash}
\bibfield{author}{\bibinfo{person}{Binbin Hu}, \bibinfo{person}{Zhiqiang Zhang}, \bibinfo{person}{Chuan Shi}, \bibinfo{person}{Jun Zhou}, \bibinfo{person}{Xiaolong Li}, {and} \bibinfo{person}{Yuan Qi}.} \bibinfo{year}{2019}\natexlab{}.
\newblock \showarticletitle{{Cash-out User Detection Based on Attributed Heterogeneous Information Network with a Hierarchical Attention Mechanism}}. In \bibinfo{booktitle}{\emph{Proceedings of the AAAI Conference on Artificial Intelligence}}.
\newblock


\bibitem[Hu et~al\mbox{.}(2021)]%
        {hu2021ogblsc}
\bibfield{author}{\bibinfo{person}{Weihua Hu}, \bibinfo{person}{Matthias Fey}, \bibinfo{person}{Hongyu Ren}, \bibinfo{person}{Maho Nakata}, \bibinfo{person}{Yuxiao Dong}, {and} \bibinfo{person}{Jure Leskovec}.} \bibinfo{year}{2021}\natexlab{}.
\newblock \showarticletitle{{OGB-LSC: A Large-Scale Challenge for Machine Learning on Graphs}}. In \bibinfo{booktitle}{\emph{Proceedings of the Neural Information Processing Systems Track on Datasets and Benchmarks}}.
\newblock


\bibitem[Hu et~al\mbox{.}(2020b)]%
        {hu2020open}
\bibfield{author}{\bibinfo{person}{Weihua Hu}, \bibinfo{person}{Matthias Fey}, \bibinfo{person}{Marinka Zitnik}, \bibinfo{person}{Yuxiao Dong}, \bibinfo{person}{Hongyu Ren}, \bibinfo{person}{Bowen Liu}, \bibinfo{person}{Michele Catasta}, {and} \bibinfo{person}{Jure Leskovec}.} \bibinfo{year}{2020}\natexlab{b}.
\newblock \showarticletitle{{Open Graph Benchmark: Datasets for Machine Learning on Graphs}}. In \bibinfo{booktitle}{\emph{Proceedins of Advances in Neural Information Processing systems}}.
\newblock


\bibitem[Hu et~al\mbox{.}(2020a)]%
        {hu2020heterogeneous}
\bibfield{author}{\bibinfo{person}{Ziniu Hu}, \bibinfo{person}{Yuxiao Dong}, \bibinfo{person}{Kuansan Wang}, {and} \bibinfo{person}{Yizhou Sun}.} \bibinfo{year}{2020}\natexlab{a}.
\newblock \showarticletitle{{Heterogeneous Graph Transformer}}. In \bibinfo{booktitle}{\emph{Proceedings of the World Web Conference}}.
\newblock


\bibitem[Jia et~al\mbox{.}(2020)]%
        {jia2020roc}
\bibfield{author}{\bibinfo{person}{Zhihao Jia}, \bibinfo{person}{Sina Lin}, \bibinfo{person}{Mingyu Gao}, \bibinfo{person}{Matei Zaharia}, {and} \bibinfo{person}{Alex Aiken}.} \bibinfo{year}{2020}\natexlab{}.
\newblock \showarticletitle{{Improving the Accuracy, Scalability, and Performance of Graph Neural Networks with ROC}}. In \bibinfo{booktitle}{\emph{Proceedings of Machine Learning and Systems}}.
\newblock


\bibitem[Kaggle(2024)]%
        {donor}
\bibfield{author}{\bibinfo{person}{Kaggle}.} \bibinfo{year}{2024}\natexlab{}.
\newblock \bibinfo{title}{DonorsChoose.org Application Screening}.
\newblock \bibinfo{howpublished}{\url{https://www.kaggle.com/competitions/donorschoose-application-screening/data}}.
\newblock


\bibitem[Kalantzi and Karypis(2021)]%
        {kalantzi2021position}
\bibfield{author}{\bibinfo{person}{Maria Kalantzi} {and} \bibinfo{person}{George Karypis}.} \bibinfo{year}{2021}\natexlab{}.
\newblock \showarticletitle{{Position-based Hash Embeddings for Scaling Graph Neural Networks}}. In \bibinfo{booktitle}{\emph{2021 IEEE International Conference on Big Data}}.
\newblock


\bibitem[Kaler et~al\mbox{.}(2023)]%
        {kaler2023communication}
\bibfield{author}{\bibinfo{person}{Tim Kaler}, \bibinfo{person}{Alexandros Iliopoulos}, \bibinfo{person}{Philip Murzynowski}, \bibinfo{person}{Tao Schardl}, \bibinfo{person}{Charles~E Leiserson}, {and} \bibinfo{person}{Jie Chen}.} \bibinfo{year}{2023}\natexlab{}.
\newblock \showarticletitle{{Communication-Efficient Graph Neural Networks with Probabilistic Neighborhood Expansion Analysis and Caching}}. In \bibinfo{booktitle}{\emph{Proceedings of Machine Learning and Systems}}.
\newblock


\bibitem[Karypis and Kumar(1998)]%
        {metis1998}
\bibfield{author}{\bibinfo{person}{George Karypis} {and} \bibinfo{person}{Vipin Kumar}.} \bibinfo{year}{1998}\natexlab{}.
\newblock \showarticletitle{{A Fast and High Quality Multilevel Scheme for Partitioning Irregular Graphs}}.
\newblock \bibinfo{journal}{\emph{SIAM Journal on scientific Computing}} (\bibinfo{year}{1998}).
\newblock


\bibitem[Khatua et~al\mbox{.}(2023)]%
        {khatua2023igb}
\bibfield{author}{\bibinfo{person}{Arpandeep Khatua}, \bibinfo{person}{Vikram~Sharma Mailthody}, \bibinfo{person}{Bhagyashree Taleka}, \bibinfo{person}{Tengfei Ma}, \bibinfo{person}{Xiang Song}, {and} \bibinfo{person}{Wen-mei Hwu}.} \bibinfo{year}{2023}\natexlab{}.
\newblock \showarticletitle{{IGB: Addressing The Gaps In Labeling, Features, Heterogeneity, and Size of Public Graph Datasets for Deep Learning Research}}. In \bibinfo{booktitle}{\emph{Proceedings of the 29th ACM SIGKDD Conference on Knowledge Discovery and Data Mining}}.
\newblock


\bibitem[Kingma and Ba(2014)]%
        {kingma2014adam}
\bibfield{author}{\bibinfo{person}{Diederik~P Kingma} {and} \bibinfo{person}{Jimmy Ba}.} \bibinfo{year}{2014}\natexlab{}.
\newblock \showarticletitle{{Adam: A Method for Stochastic Optimization}}.
\newblock \bibinfo{journal}{\emph{arXiv preprint}} (\bibinfo{year}{2014}).
\newblock


\bibitem[Kipf and Welling(2017)]%
        {kipf2017semisupervised}
\bibfield{author}{\bibinfo{person}{Thomas~N. Kipf} {and} \bibinfo{person}{Max Welling}.} \bibinfo{year}{2017}\natexlab{}.
\newblock \showarticletitle{{Semi-Supervised Classification with Graph Convolutional Networks}}. In \bibinfo{booktitle}{\emph{Proceedings of International Conference on Learning Representations}}.
\newblock


\bibitem[Li et~al\mbox{.}(2020)]%
        {li13pytorch}
\bibfield{author}{\bibinfo{person}{Shen Li}, \bibinfo{person}{Yanli Zhao}, \bibinfo{person}{Rohan Varma}, \bibinfo{person}{Omkar Salpekar}, \bibinfo{person}{Pieter Noordhuis}, \bibinfo{person}{Teng Li}, \bibinfo{person}{Adam Paszke}, \bibinfo{person}{Jeff Smith}, \bibinfo{person}{Brian Vaughan}, \bibinfo{person}{Pritam Damania}, {et~al\mbox{.}}} \bibinfo{year}{2020}\natexlab{}.
\newblock \showarticletitle{{PyTorch Distributed: Experiences on Accelerating Data Parallel Training}}. In \bibinfo{booktitle}{\emph{Proceedings of the VLDB Endowment}}.
\newblock


\bibitem[Lin et~al\mbox{.}(2020)]%
        {lin2020pagraph}
\bibfield{author}{\bibinfo{person}{Zhiqi Lin}, \bibinfo{person}{Cheng Li}, \bibinfo{person}{Youshan Miao}, \bibinfo{person}{Yunxin Liu}, {and} \bibinfo{person}{Yinlong Xu}.} \bibinfo{year}{2020}\natexlab{}.
\newblock \showarticletitle{{PaGraph: Scaling GNN Training on Large Graphs via Computation-aware Caching}}. In \bibinfo{booktitle}{\emph{Proceedings of the 11th ACM Symposium on Cloud Computing}}.
\newblock


\bibitem[Liu et~al\mbox{.}(2023)]%
        {liu2021bgl}
\bibfield{author}{\bibinfo{person}{Tianfeng Liu}, \bibinfo{person}{Yangrui Chen}, \bibinfo{person}{Dan Li}, \bibinfo{person}{Chuan Wu}, \bibinfo{person}{Yibo Zhu}, \bibinfo{person}{Jun He}, \bibinfo{person}{Yanghua Peng}, \bibinfo{person}{Hongzheng Chen}, \bibinfo{person}{Hongzhi Chen}, {and} \bibinfo{person}{Chuanxiong Guo}.} \bibinfo{year}{2023}\natexlab{}.
\newblock \showarticletitle{{BGL: GPU-efficient GNN Training by Optimizing Graph Data I/O and Preprocessing}}. In \bibinfo{booktitle}{\emph{Proceedings of the 20th USENIX Symposium on Networked Systems Design and Implementation}}.
\newblock


\bibitem[Malewicz et~al\mbox{.}(2010)]%
        {malewicz2010pregel}
\bibfield{author}{\bibinfo{person}{Grzegorz Malewicz}, \bibinfo{person}{Matthew~H Austern}, \bibinfo{person}{Aart~JC Bik}, \bibinfo{person}{James~C Dehnert}, \bibinfo{person}{Ilan Horn}, \bibinfo{person}{Naty Leiser}, {and} \bibinfo{person}{Grzegorz Czajkowski}.} \bibinfo{year}{2010}\natexlab{}.
\newblock \showarticletitle{{Pregel: A System for Large-scale Graph Processing}}. In \bibinfo{booktitle}{\emph{Proceedings of the 2010 ACM SIGMOD International Conference on Management of Data}}.
\newblock


\bibitem[Ni et~al\mbox{.}(2019)]%
        {ni2019justifying}
\bibfield{author}{\bibinfo{person}{Jianmo Ni}, \bibinfo{person}{Jiacheng Li}, {and} \bibinfo{person}{Julian McAuley}.} \bibinfo{year}{2019}\natexlab{}.
\newblock \showarticletitle{{Justifying Recommendations Using Distantly-labeled Reviews and Fine-grained Aspects}}. In \bibinfo{booktitle}{\emph{{Proceedings of the 2019 Conference on Empirical Methods in Natural Language Processing and the 9th International Joint Conference on Natural Language Processing}}}.
\newblock


\bibitem[NVIDIA(2023a)]%
        {cudap2p}
\bibfield{author}{\bibinfo{person}{NVIDIA}.} \bibinfo{year}{2023}\natexlab{a}.
\newblock \bibinfo{title}{{CUDA Peer Device Memory Access}}.
\newblock
\newblock
\newblock
\shownote{\url{https://docs.nvidia.com/cuda/cuda-runtime-api/group__CUDART__PEER.html}}.


\bibitem[NVIDIA(2023b)]%
        {nccl}
\bibfield{author}{\bibinfo{person}{NVIDIA}.} \bibinfo{year}{2023}\natexlab{b}.
\newblock \bibinfo{title}{{NCCL}}.
\newblock
\newblock
\newblock
\shownote{\url{https://github.com/NVIDIA/nccl}}.


\bibitem[Paszke et~al\mbox{.}(2019)]%
        {paszke2019pytorch}
\bibfield{author}{\bibinfo{person}{Adam Paszke}, \bibinfo{person}{Sam Gross}, \bibinfo{person}{Francisco Massa}, \bibinfo{person}{Adam Lerer}, \bibinfo{person}{James Bradbury}, \bibinfo{person}{Gregory Chanan}, \bibinfo{person}{Trevor Killeen}, \bibinfo{person}{Zeming Lin}, \bibinfo{person}{Natalia Gimelshein}, \bibinfo{person}{Luca Antiga}, {et~al\mbox{.}}} \bibinfo{year}{2019}\natexlab{}.
\newblock \showarticletitle{{PyTorch: An Imperative Style, High-performance Deep Learning Library}}. In \bibinfo{booktitle}{\emph{Proceedings of Advances in Neural Information Processing Systems}}.
\newblock


\bibitem[Petroni et~al\mbox{.}(2015)]%
        {petroni2015hdrf}
\bibfield{author}{\bibinfo{person}{Fabio Petroni}, \bibinfo{person}{Leonardo Querzoni}, \bibinfo{person}{Khuzaima Daudjee}, \bibinfo{person}{Shahin Kamali}, {and} \bibinfo{person}{Giorgio Iacoboni}.} \bibinfo{year}{2015}\natexlab{}.
\newblock \showarticletitle{{Hdrf: Stream-based Partitioning for Power-law Graphs}}. In \bibinfo{booktitle}{\emph{Proceedings of the 24th ACM international on conference on information and knowledge management}}.
\newblock


\bibitem[Schlichtkrull et~al\mbox{.}(2018)]%
        {schlichtkrull2018modeling}
\bibfield{author}{\bibinfo{person}{Michael Schlichtkrull}, \bibinfo{person}{Thomas~N Kipf}, \bibinfo{person}{Peter Bloem}, \bibinfo{person}{Rianne Van Den~Berg}, \bibinfo{person}{Ivan Titov}, {and} \bibinfo{person}{Max Welling}.} \bibinfo{year}{2018}\natexlab{}.
\newblock \showarticletitle{{Modeling Relational Data with Graph Convolutional Networks}}. In \bibinfo{booktitle}{\emph{Proceedings of The Semantic Web: 15th International Conference}}.
\newblock


\bibitem[Sun et~al\mbox{.}(2023)]%
        {sun2023legion}
\bibfield{author}{\bibinfo{person}{Jie Sun}, \bibinfo{person}{Li Su}, \bibinfo{person}{Zuocheng Shi}, \bibinfo{person}{Wenting Shen}, \bibinfo{person}{Zeke Wang}, \bibinfo{person}{Lei Wang}, \bibinfo{person}{Jie Zhang}, \bibinfo{person}{Yong Li}, \bibinfo{person}{Wenyuan Yu}, \bibinfo{person}{Jingren Zhou}, {and} \bibinfo{person}{Fei Wu}.} \bibinfo{year}{2023}\natexlab{}.
\newblock \showarticletitle{{Legion: Automatically Pushing the Envelope of {Multi-GPU} System for {Billion-Scale} {GNN} Training}}. In \bibinfo{booktitle}{\emph{Proceedings of 2023 USENIX Annual Technical Conference}}.
\newblock


\bibitem[Team(2023)]%
        {graphlearn}
\bibfield{author}{\bibinfo{person}{GLT Team}.} \bibinfo{year}{2023}\natexlab{}.
\newblock \bibinfo{title}{{GraphLearn for PyTorch}}.
\newblock
\newblock
\newblock
\shownote{\url{https://github.com/alibaba/graphlearn-for-pytorch}}.


\bibitem[Veličković et~al\mbox{.}(2018)]%
        {velickovic2018graph}
\bibfield{author}{\bibinfo{person}{Petar Veličković}, \bibinfo{person}{Guillem Cucurull}, \bibinfo{person}{Arantxa Casanova}, \bibinfo{person}{Adriana Romero}, \bibinfo{person}{Pietro Liò}, {and} \bibinfo{person}{Yoshua Bengio}.} \bibinfo{year}{2018}\natexlab{}.
\newblock \showarticletitle{{Graph Attention Networks}}. In \bibinfo{booktitle}{\emph{Proceedings of International Conference on Learning Representations}}.
\newblock


\bibitem[Wan et~al\mbox{.}(2023)]%
        {wan2023adaptive}
\bibfield{author}{\bibinfo{person}{Borui Wan}, \bibinfo{person}{Juntao Zhao}, {and} \bibinfo{person}{Chuan Wu}.} \bibinfo{year}{2023}\natexlab{}.
\newblock \showarticletitle{{Adaptive Message Quantization and Parallelization for Distributed Full-Graph GNN Training}}. In \bibinfo{booktitle}{\emph{Proceedings of Machine Learning and Systems}}.
\newblock


\bibitem[Wan et~al\mbox{.}(2022)]%
        {wan2022bns}
\bibfield{author}{\bibinfo{person}{Cheng Wan}, \bibinfo{person}{Youjie Li}, \bibinfo{person}{Ang Li}, \bibinfo{person}{Nam~Sung Kim}, {and} \bibinfo{person}{Yingyan Lin}.} \bibinfo{year}{2022}\natexlab{}.
\newblock \showarticletitle{{BNS-GCN: Efficient Full-graph Training of Graph Convolutional Networks with Partition-parallelism and Random Boundary Node Sampling}}. In \bibinfo{booktitle}{\emph{Proceedings of Machine Learning and Systems}}.
\newblock


\bibitem[Wan et~al\mbox{.}(2021)]%
        {wan2021pipegcn}
\bibfield{author}{\bibinfo{person}{Cheng Wan}, \bibinfo{person}{Youjie Li}, \bibinfo{person}{Cameron~R Wolfe}, \bibinfo{person}{Anastasios Kyrillidis}, \bibinfo{person}{Nam~Sung Kim}, {and} \bibinfo{person}{Yingyan Lin}.} \bibinfo{year}{2021}\natexlab{}.
\newblock \showarticletitle{{PipeGCN: Efficient Full-Graph Training of Graph Convolutional Networks with Pipelined Feature Communication}}. In \bibinfo{booktitle}{\emph{Proceedings of International Conference on Learning Representations}}.
\newblock


\bibitem[Wang et~al\mbox{.}(2020)]%
        {wang2020microsoft}
\bibfield{author}{\bibinfo{person}{Kuansan Wang}, \bibinfo{person}{Zhihong Shen}, \bibinfo{person}{Chiyuan Huang}, \bibinfo{person}{Chieh-Han Wu}, \bibinfo{person}{Yuxiao Dong}, {and} \bibinfo{person}{Anshul Kanakia}.} \bibinfo{year}{2020}\natexlab{}.
\newblock \showarticletitle{{Microsoft Academic Graph: When Experts are Not Enough}}.
\newblock \bibinfo{journal}{\emph{Quantitative Science Studies}} (\bibinfo{year}{2020}).
\newblock


\bibitem[Wang et~al\mbox{.}(2022)]%
        {wang2022survey}
\bibfield{author}{\bibinfo{person}{Xiao Wang}, \bibinfo{person}{Deyu Bo}, \bibinfo{person}{Chuan Shi}, \bibinfo{person}{Shaohua Fan}, \bibinfo{person}{Yanfang Ye}, {and} \bibinfo{person}{S~Yu Philip}.} \bibinfo{year}{2022}\natexlab{}.
\newblock \showarticletitle{{A Survey on Heterogeneous Graph Embedding: Methods, Techniques, Applications and Sources}}.
\newblock \bibinfo{journal}{\emph{IEEE Transactions on Big Data}} (\bibinfo{year}{2022}).
\newblock


\bibitem[Wang et~al\mbox{.}(2019)]%
        {wang2019heterogeneous}
\bibfield{author}{\bibinfo{person}{Xiao Wang}, \bibinfo{person}{Houye Ji}, \bibinfo{person}{Chuan Shi}, \bibinfo{person}{Bai Wang}, \bibinfo{person}{Yanfang Ye}, \bibinfo{person}{Peng Cui}, {and} \bibinfo{person}{Philip~S Yu}.} \bibinfo{year}{2019}\natexlab{}.
\newblock \showarticletitle{{Heterogeneous Graph Attention Network}}. In \bibinfo{booktitle}{\emph{Proceedings of the world wide web conference}}.
\newblock


\bibitem[Xie et~al\mbox{.}(2022)]%
        {xie2022fleche}
\bibfield{author}{\bibinfo{person}{Minhui Xie}, \bibinfo{person}{Youyou Lu}, \bibinfo{person}{Jiazhen Lin}, \bibinfo{person}{Qing Wang}, \bibinfo{person}{Jian Gao}, \bibinfo{person}{Kai Ren}, {and} \bibinfo{person}{Jiwu Shu}.} \bibinfo{year}{2022}\natexlab{}.
\newblock \showarticletitle{{Fleche: An Efficient GPU Embedding Cache for Personalized Recommendations}}. In \bibinfo{booktitle}{\emph{Proceedings of the Seventeenth European Conference on Computer Systems}}.
\newblock


\bibitem[Yang et~al\mbox{.}(2022)]%
        {yang2022gnnlab}
\bibfield{author}{\bibinfo{person}{Jianbang Yang}, \bibinfo{person}{Dahai Tang}, \bibinfo{person}{Xiaoniu Song}, \bibinfo{person}{Lei Wang}, \bibinfo{person}{Qiang Yin}, \bibinfo{person}{Rong Chen}, \bibinfo{person}{Wenyuan Yu}, {and} \bibinfo{person}{Jingren Zhou}.} \bibinfo{year}{2022}\natexlab{}.
\newblock \showarticletitle{{GNNLab: A Factored System for Sample-based GNN Training over GPUs}}. In \bibinfo{booktitle}{\emph{Proceedings of the 17th European Conference on Computer Systems}}.
\newblock


\bibitem[Yin et~al\mbox{.}(2022)]%
        {yin2022nimble}
\bibfield{author}{\bibinfo{person}{Chunxing Yin}, \bibinfo{person}{Da Zheng}, \bibinfo{person}{Israt Nisa}, \bibinfo{person}{Christos Faloutsos}, \bibinfo{person}{George Karypis}, {and} \bibinfo{person}{Richard Vuduc}.} \bibinfo{year}{2022}\natexlab{}.
\newblock \showarticletitle{{Nimble GNN Embedding with Tensor-train Decomposition}}. In \bibinfo{booktitle}{\emph{Proceedings of the 28th ACM SIGKDD Conference on Knowledge Discovery and Data Mining}}.
\newblock


\bibitem[Zheng et~al\mbox{.}(2020)]%
        {zheng2020distdgl}
\bibfield{author}{\bibinfo{person}{Da Zheng}, \bibinfo{person}{Chao Ma}, \bibinfo{person}{Minjie Wang}, \bibinfo{person}{Jinjing Zhou}, \bibinfo{person}{Qidong Su}, \bibinfo{person}{Xiang Song}, \bibinfo{person}{Quan Gan}, \bibinfo{person}{Zheng Zhang}, {and} \bibinfo{person}{George Karypis}.} \bibinfo{year}{2020}\natexlab{}.
\newblock \showarticletitle{{DistDGL: Distributed Graph Neural Network Training for Billion-scale Graphs}}. In \bibinfo{booktitle}{\emph{Proceedings of 2020 IEEE/ACM 10th Workshop on Irregular Applications: Architectures and Algorithms}}.
\newblock


\bibitem[Zheng et~al\mbox{.}(2022)]%
        {zheng2022distributed}
\bibfield{author}{\bibinfo{person}{Da Zheng}, \bibinfo{person}{Xiang Song}, \bibinfo{person}{Chengru Yang}, \bibinfo{person}{Dominique LaSalle}, {and} \bibinfo{person}{George Karypis}.} \bibinfo{year}{2022}\natexlab{}.
\newblock \showarticletitle{{Distributed Hybrid CPU and GPU Training for Graph Neural Networks on Billion-scale Heterogeneous Graphs}}. In \bibinfo{booktitle}{\emph{Proceedings of the 28th ACM SIGKDD Conference on Knowledge Discovery and Data Mining}}.
\newblock


\bibitem[Zheng et~al\mbox{.}(2021)]%
        {zheng2021multi}
\bibfield{author}{\bibinfo{person}{Jiawei Zheng}, \bibinfo{person}{Qianli Ma}, \bibinfo{person}{Hao Gu}, {and} \bibinfo{person}{Zhenjing Zheng}.} \bibinfo{year}{2021}\natexlab{}.
\newblock \showarticletitle{{Multi-view Denoising Graph Auto-encoders on Heterogeneous Information Networks for Cold-start Recommendation}}. In \bibinfo{booktitle}{\emph{Proceedings of the 27th ACM SIGKDD Conference on Knowledge Discovery \& Data Mining}}.
\newblock


\bibitem[Zhong et~al\mbox{.}(2023)]%
        {zhong2023gnnflow}
\bibfield{author}{\bibinfo{person}{Yuchen Zhong}, \bibinfo{person}{Guangming Sheng}, \bibinfo{person}{Tianzuo Qin}, \bibinfo{person}{Minjie Wang}, \bibinfo{person}{Quan Gan}, {and} \bibinfo{person}{Chuan Wu}.} \bibinfo{year}{2023}\natexlab{}.
\newblock \showarticletitle{{GNNFlow: A Distributed Framework for Continuous Temporal GNN Learning on Dynamic Graphs}}.
\newblock \bibinfo{journal}{\emph{arXiv preprint}} (\bibinfo{year}{2023}).
\newblock


\end{thebibliography}

\clearpage
\begin{proof}[Proposition~\ref{prop:equivalence}]
    Let $v$ be an arbitrary target vertex. In the vanilla scheme, the embedding of $\mathbf{h}_v^{\mathrm{vanilla}}$ is computed as,
    {\small
    \begin{align}
        \mathbf{h}_v^{(l)} = \mathrm{AGG}^{(l)}_{\mathrm{all}}\left(\left\{ \mathrm{AGG}^{(l)}_{r}\left(\left\{\mathbf{h}^{(l-1)}_{u}, u\in N_{r}(v)\right\} \right), r \in \mathcal{R} \right\} \right)
    \end{align}
    }
    
    Under RAF, the embedding of $\mathbf{h}_v^{\mathrm{RAF}}$  is computed through two steps.
    
    First, by lines 2-6 in RAF.~\ref{alg:raf_paradigm}, the intermediate embedding of the vertex from different relations is computed remotely and aggregated at the remote worker. This yields a set of partial aggregations,
    $$\left\{ \mathrm{AGG}^{(l)}_{r}\color{black}\left(\left\{\mathbf{h}^{(l-1)}_{u}, u\in N_{r}(v)\right\} \right), r \in \mathcal{R} \right\}$$
    Then, these messages are sent to the worker with the target vertex, and the worker performs cross-relation aggregation $\mathrm{AGG}^{(l)}_{\mathrm{all}}$ on these partial aggregations (lines 8-12 in RAF.~\ref{alg:raf_paradigm}).

    We have that $$\mathbf{h}_v^{\mathrm{RAF}} = \mathbf{h}_v^{\mathrm{vanilla}}$$
\end{proof}
\begin{proof}[Proposition~\ref{prop:communication_complexity}]
    Given an arbitrary partition $G_1, G_2$, we first focus our analysis on $G_1$. Let $k$ denote the number of relations in the graphs, and recall that $\mathrm{B}(G_1)$ is the number of boundary vertex in $G_1$.

    Let $v$ be an arbitrary vertex in $G_1$ that requires communication from $G_2$. By definition, $v$ must be a boundary vertex as it has a neighborhood from partition $G_2$.

    By lines 4-12 in RAF.~\ref{alg:raf_paradigm}, only the intermediate representation of each relation $r$ in $v$'s neighborhood needs to be communicated. Therefore, the number of messages received for $v$ in $G_1$ is at most $k$.

    Therefore, the number of messages received for $G_1$ from $G_2$ is at most $$k \mathrm{B}(G_1).$$
    Since $k$ is a constant for a given dataset or task, we have the communication complexity of worker $G_1$ is $$\Theta(\mathrm{B}(G_1)).$$ 

    This completes the proof since the analysis for $G_2$ is symmetric.

\end{proof}
\begin{proof}[Proposition~\ref{prop:communication_complexity2}]
    Again, let us focus on the analysis in $G_1$. 
    
    By definition of boundary vertex, a boundary vertex in $G_1$ must be an endpoint of at least one cross-partition edge. However, there may be multiple cross-partition edges connecting to the vertex. Therefore, we have that
    $$\mathrm{B}(G_1) \leq \mathrm{E}(G_1,G_2).$$

    By a symmetric argument, we have that
    $$\mathrm{B}(G_2) \leq \mathrm{E}(G_1,G_2).$$

    Combining the result above, we can obtain that 
    $$\max\{\mathrm{B}(G_1),\mathrm{B}(G_2) \} \leq \mathrm{E}(G_1,G_2).$$
\end{proof}

\end{document}